\documentclass[11pt]{article}
\usepackage{amssymb,amscd,array}
\catcode `\@=11
\@addtoreset{equation}{subsection}

\def\qed{\blacksquare}
\newcommand{\be}{\begin{equation}}
\newcommand{\ee}{\end{equation}}
\newcommand{\bea}{\begin{eqnarray}}
\newcommand{\eea}{\end{eqnarray}}
\newcommand{\R}{\mathbb{R}}
\newcommand{\N}{\mathbb{N}}
\newcommand{\C}{\mathbb{C}}
\newtheorem{thm}{Theorem}[section]
\newtheorem{rem}[thm]{Remark}
\newtheorem{lemma}[thm]{Lemma}

\newtheorem{prop}[thm]{Proposition}
\begin{document}
\begin{titlepage}
%\thispagestyle{empty}
%\begin{flushright}
%IFA-FT-402-1994, November
%\end{flushright}
%\bigskip\bigskip
\begin{center}
{\bf \Large{Scale Invariance in the Causal Approach to 
Renormalization Theory\\}}
\end{center}
\vskip 1.0truecm
\centerline{D. R. Grigore
\footnote{e-mail: grigore@theor1.theory.nipne.ro, grigore@theory.nipne.ro}}
\vskip5mm
\centerline{Dept. of Theor. Phys., Inst. Atomic Phys.}
\centerline{Bucharest-M\u agurele, P. O. Box MG 6, ROM\^ANIA}
\vskip 2cm
\bigskip \nopagebreak
\begin{abstract}
\noindent
The dilation invariance is studied in the framework of Epstein-Glaser approach
to renormalization theory. Some analogues of the Callan-Symanzik equations are
found and they are applied to the scalar field theory and to Yang-Mills models.
We find the interesting result that, if all the fields of the theory
have zero masses, then from purely cohomological consideration, one
can obtain the anomalous terms of logarithmic type.  
\end{abstract}
%\newpage\setcounter{page}1
\end{titlepage}

\section{Introduction}

The causal approach to renormalization theory of by Epstein and Glaser
\cite{EG1}, \cite{Gl} leads to important simplification of the renormalization
theory as well as of the computational aspects. This approach works for
quantum electrodynamics \cite{Sc1}, Yang-Mills theories \cite{DHKS1} 
\cite{DHKS2} \cite{DHS2} \cite{DHS3} \cite{AS} \cite{ASD3} \cite{D1}, 
\cite{D2}, \cite{Hu1}-\cite{Hu4}, \cite {Kr1} \cite{Kr3} \cite{DS}, gravitation
\cite{Gri1}, \cite{Gri2}, \cite{SW}, etc.

In this paper we investigate the r\^ole of dilation invariance in the causal
approach. 
The pioneering works on scale covariance in perturbative field theory
are \cite{Ca}, \cite{CCJ} and \cite{CJ}. A mathematical refined analysis was
developed in \cite{Si1} and \cite{Si2}, the main mathematical tool being the
so-called quantum action principle \cite{Lo} (for a review see \cite{PR}).

Our strategy will be based exclusively on the Epstein-Glaser
construction of the chonological producs for the free fields.
In the next Section we define the dilation invariance operator for 
various free fields. Next, we remind the basic facts about renormalization
theory. We will emphasize the original Epstein-Glaser approach where one
considers a set of (linearly independent) interaction Lagrangian and attaches
to each of this Lagrangian a (space-time dependent) coupling constant. Then we
are able to prove the basic theorem concerning the arbitrariness of the
chronological products for the same set of interaction Lagrangian. This
problem was already addressed in \cite{BS}, \cite{Pi}, but we argue
that the natural framework is the multi-valued coupling constant
approach of \cite{EG1}.

In Section 4 we obtain consequences about the scale behaviour of the
chronological products. One expects that the action of the dilations
operators on the chronological product should give the usual scaled
chronological products, up to some {\it scale anomalies}. We will
prove that one can reduce the analysis
to a cohomological problem. An important difference appears in the
study of this problem in the cases of a massless and of a massive field. 
If there exists at least one massive field in the theory then we can
prove that one can choose the chronological product to be scale
covariant. In the opposite case, one finds out that some anoumalous
terms of logarithmic behaviour can appear. We emphasis again that these
results hold for the chronological products of the free fields. We
will comment on what one should expect for the case of the interacting
fields in the last Section.

We will apply these considerations for Yang-Mills models in Section
\ref{ym} and obtain a
restrictions on the possible form of the anomalies, namely the
canonical dimension of such an anomalous expression must be $5$.

\newpage

\section{Dilation Invariance in Quantum Field Theory\label{dilation-qft}}

It is well known that the Fock space of the real scalar field of mass $m$ can
be defined as:
\be
{\cal F}_{m} \equiv \oplus_{n=0}^{\infty} {\cal F}^{(n)}_{m}
\ee
where
${\cal F}^{(n)}_{m}$
is the set of Borel function
$\Phi^{(n)}: (X_{m}^{+})^{\otimes n} \rightarrow \C$
which are square integrable with respect to the Lorentz invariant measure:
$d\alpha_{m}^{+}(p) \equiv {d{\bf p}\over \sqrt{{\bf p}^{2} + m^{2}}}$
and completely symmetric in the all variables (see \cite{Va} for notations). 
Then we have:
\begin{prop}
Let us define for any
$\lambda \in \R_{+}$
the operators 
$
{\cal U}_{\lambda}: {\cal F}_{m}
\longrightarrow {\cal F}_{\lambda^{-1}m}
$
as follows:
\be
\left({\cal U}_{\lambda}\Phi^{(n)}\right)(p_{1},\dots,p_{n}) =
\lambda^{n} \quad
\Phi^{(n)}(\lambda p_{1},\dots,\lambda p_{n}).
\label{u-lambda}
\ee

Then:

(i) The operators 
${\cal U}_{\lambda}$
are unitary;

(ii) The following relations are verified for all 
$\lambda, \lambda' \in \R_{+}$:
\be
{\cal U}_{\lambda} {\cal U}_{\lambda'} = {\cal U}_{\lambda\lambda'};
\ee

(iii) If 
${\cal U}^{[m]}_{a,L}$
is the representation of the Poincar\'e group in the Fock space
${\cal F}^{(n)}_{m}$,
then:
\be
{\cal U}^{[m]}_{a,L} \quad {\cal U}_{\lambda} =
{\cal U}_{\lambda} \quad {\cal U}^{[\lambda m]}_{\lambda^{-1}a,L}
\label{reps}
\ee
for all translations $a$ and all Lorentz transformations $L$.
\label{dilation}
\end{prop}

{\bf Proof:}
The proof of the first assertion is based on the scaling properties of the
measure 
$d\alpha_{m}^{+}(p)$.
The next assertions follow from elementary computations.
$\qed$

If we use the definition of the annihilation operators
\be
\left(a(q;m)\Phi\right)^{(n)}(p_{1},\dots,p_{n}) =
\sqrt{n + 1} \quad \Phi^{(n+1)}(q,p_{1},\dots,p_{n}) 
\ee
then we immediately get the identity:
\be
{\cal U}_{\lambda} \quad a(q;m) \quad {\cal U}_{\lambda}^{-1} = 
\lambda \quad a(\lambda^{-1} q; \lambda^{-1} m).
\label{scale-anihilator}
\ee

By hermitian conjugation we get a similar identity for the creation operators
$a^{*}(q)$.

The expression of the real scalar field of mass $m$ is:
\be
\phi(x;m) \equiv {1\over (2\pi)^{3/2}} \int d\alpha_{m}^{+}(p)
\left[ e^{-i x\cdot p} a(p;m) + e^{i x\cdot p} a^{*}(p;m) \right]
\label{scalar}
\ee
so we get from (\ref{scale-anihilator}) the following relation:
\be
{\cal U}_{\lambda} \quad \phi(x;m) \quad {\cal U}_{\lambda}^{-1} =
\lambda \phi(\lambda x; \lambda^{-1} m).
\label{scale-phi}
\ee

\begin{rem}
There is an alternative point of view. One can define the operators
$
{\cal U}'_{\lambda}: {\cal F}_{m} \longrightarrow {\cal F}_{m}
$
according to
\be
\left({\cal U}'_{\lambda}\Phi^{(n)}\right)({\bf p}_{1},\dots,{\bf p}_{n}) =
\prod_{i=1}^{n}  r_{\lambda}({\bf p_{i}}) \quad
\Phi^{(n)}(\lambda {\bf p}_{1},\dots,\lambda {\bf p}_{n})
\ee
where
\be
r_{\lambda}({\bf p}) \equiv \lambda^{3/2} 
\sqrt{\omega_{m}({\bf p})\over \omega_{m}(\lambda {\bf p})}.
\ee

Because we have the cocycle identity
\be
r_{\lambda}({\bf p}) \quad r_{\lambda'}(\lambda {\bf p}) =
r_{\lambda \lambda'}({\bf p})
\ee
the map 
$\lambda \rightarrow {\cal U}'_{\lambda}$
defined above is a representation of the multiplicative group 
$\R_{+}$
(the {\it dilation}) group in the Fock space of the scalar field. Moreover, the
relations (\ref{reps}), (\ref{scale-anihilator}) and (\ref{scale-phi}) are
valid only up terms of order $O(m)$ because we have
$
r_{\lambda}({\bf p}) = \lambda + O(m).
$
So, we see that some information is lost in this approach.
\end{rem}

It is easy to prove that relations of the same type as (\ref{scale-phi}) are
valid for other types of fields, namely fields of integer spin. This includes
the electromagnetic potential, the Yang-Mills fields, the gravitational field
and also the ghosts fields used in the process of quantization. For a Dirac
field an important difference appears. Instead of (\ref{scalar}) we have:
\be
\psi(x;M) \equiv {1\over (2\pi)^{3/2}} \int d\alpha_{M}^{+}(p)
\left[ e^{-i x\cdot p} \sum_{i=1}^{2} u_{i}(p;M) b_{i}(p;M) +
e^{i x\cdot p} \sum_{i=1}^{2} v_{i}(p;M) b_{i}^{*}(p;M) \right]
\label{dirac}
\ee
(see \cite{Sc1}) where 
$b^{\#}_{i}(p;M)$
are the creation (annihilation) operators; the expressions
$u_{i}(p;M)$
and
$v_{i}(p;M)$
are solutions of the free Dirac equation of positive (negative) values. To have
Poincar\'e covariance of the field operator 
$\psi$
one has to normalize in such a way these spinors such that we have:
\be
u_{i}(\lambda p; \lambda M) = \lambda^{1/2} u_{i}(p;M), \quad
v_{i}(\lambda p; \lambda M) = \lambda^{1/2} v_{i}(p;M).
\ee
So we get instead of (\ref{scale-phi}):
\be
{\cal U}_{\lambda} \quad \psi(x;M) \quad {\cal U}_{\lambda}^{-1} =
\lambda^{3/2} \psi(\lambda x; \lambda^{-1} M).
\label{scale-psi}
\ee

We can obviously prove that the relations (\ref{reps}) are valid in the most
general case, with fields of various spins. 

Let us note that if we apply to the relations (\ref{scale-phi}) or
(\ref{scale-psi}) a derivation operator
${\partial\over \partial x^{\mu}}$
we obtain a supplementary factor
$\lambda$
in the right hand side.

Finally, if
$W(x;{\bf m})$
is a Wick monomial in free fields of various masses
${\bf m} = (m_{1},\dots,M_{1},\dots)$
we obtain a generalization of the relations
(\ref{scale-phi}) and (\ref{scale-psi}), namely:
\be
{\cal U}_{\lambda} \quad W(x;{\bf m}) \quad {\cal U}_{\lambda}^{-1} =
\lambda^{\omega(W)} W(\lambda x;\lambda^{-1} {\bf m})
\label{scale}
\ee
where the number
$\omega(W)$
is called the {\it canonical dimension} of the monomial $W$ and is computed
according to the well known rule: one attributes to every integer (resp.
half-integer) spin field the canonical dimension $1$ (resp. $3/2$) and to every
derivative the canonical dimension $1$. Then one postulates that the canonical
dimension is an additive function. 

One can extend these considerations to Wick monomials in many variables
$W(x_{1},\dots,x_{n})$.
If the interaction Lagrangian of a model verifies a relation of the type
(\ref{scale}) we say that the model is {\it dilation} (or {\it scale})-{\it
covariant}.  It also well known that the canonical dimension of fields is an
important property in renormalization theory.

\newpage

\section{Renormalization Theory\label{ren}}

\subsection{Bogoliubov Axioms\label{bogoliubov}}

We outline here the axioms of a {\it multi-Lagrangian} perturbation theory.
Following Bogoliubov and Shirkov ideas, in \cite{EG1} one constructs the 
$S$-matrix as a formal series of operator valued distributions:
\be
S({\bf g})=1 + \sum_{n=1}^\infty{i^{n}\over n!}\int_{\R^{4n}} 
dx_{1}\cdots dx_{n}\, T_{j_{1},\dots,j_{n}}(x_{1},\cdots, x_{n})
g_{j_{1}}(x_{1})\cdots g_{j_{n}}(x_{n}),
\label{S}
\ee
where
${\bf g} = \left( g_{j}(x)\right)_{j = 1, \dots P}$
is a multi-valued tempered test function in the Minkowski space 
$\R^{4}$
that switches the interaction and
$T_{j_{1},\dots,j_{n}}(x_{1},\cdots, x_{n})$
are operator-valued distributions acting in the Fock space of some collection
of free fields. These operator-valued distributions are called {\it
chronological products} and verify some properties called in the following
{\it Bogoliubov axioms}. It is necessary to note that there is a canonical
projection $pr$ associating to the point $x_{i}$ the index $j_{i}$.
One starts from a set of {\it interaction Lagrangians}
$T_{j}(x), \quad j = 1,\dots, P$
and tries to construct the whole series
$T_{j_{1},\dots,j_{n}}, \quad n \geq 2$.

The interaction Lagrangians must satisfy some requirements such like Poincar\'e
invariance, hermiticity and causality. The natural candidates fulfilling these
demands is a linearly independent set of Wick polynomials operating in the 
Fock space (describing a system of weakly interacting particles).

The recursive process of constructing the chronological produces fixes the
chronological products almost uniquely. We will study this arbitrariness in 
detail later.

The physical $S$-matrix is obtained from 
$S({\bf g})$ 
taking the {\it adiabatic limit} which is , loosely speaking
the limit $ g_{j}(x) \rightarrow 1, \quad \forall j = 1, \dots, P.$ 

We give here the set of axioms imposed on the chronological products 
$T_{j_{1},\dots,j_{n}}$
following the notations of \cite{EG1}. 
\begin{itemize}
\item
First, it is clear that, without loosing generality, we can consider them 
{\it completely symmetrical} in all variables in the sense:
\be
T_{j_{\pi(1)},\dots,j_{\pi(n)}}(x_{\pi(1)},\cdots x_{\pi(p)}) = 
T_{p}(x_{1},\cdots x_{p}), \quad \forall
\pi \in {\cal P}_{p}.
\label{sym}
\ee
\item
Next, we must have {\it Poincar\'e invariance}. Because we will also consider
Dirac fields, we suppose that we have an unitary representation 
$(a, A) \mapsto {\cal U}_{a, A}$
of the group 
$inSL(2,\C)$
(the universal covering group of the proper orthochronous Poincar\'e group
${\cal P}^{\uparrow}_{+}$)
and a finite dimensional representation 
$A \mapsto S(A)$ 
of of the group 
$SL(2,\C)$
such that:
\bea
{\cal U}_{a, A} T_{j_{1},\dots,j_{n}}(x_{1},\cdots, x_{p}) 
{\cal U}^{-1}_{a, A} =
S(A^{-1})_{j_{1}k_{1}} \cdots S(A^{-1})_{j_{n}k_{n}} \times
\nonumber \\
T_{k_{1},\dots,k_{n}}(\delta(A)\cdot x_{1}+a,\cdots, \delta(A)\cdot 
x_{p}+a), \quad \forall A \in SL(2,\C), \forall a \in \R^{4}
\label{invariance}
\eea
where
$SL(2,\C) \ni A \mapsto \delta(A) \in {\cal P}^{\uparrow}_{+}$
is the covering map.  In particular, {\it translation invariance} is essential
for implementing Epstein-Glaser scheme of renormalization.

Sometimes it is possible to supplement this axiom by corresponding invariance
properties with respect to inversions (spatial and temporal) and charge
conjugation. For the standard model only the PCT invariance is available.
\item
The central axiom seems to be the requirement of {\it causality} which can be
written compactly as follows. Let us firstly introduce some standard notations.
Denote by
$
V^{+} \equiv \{x \in \R^{4} \vert \quad x^{2} > 0, \quad x_{0} > 0\}
$
and
$
V^{-} \equiv \{x \in \R^{4} \vert \quad x^{2} > 0, \quad x_{0} < 0\}
$
the upper (lower) lightcones and by
$\overline{V^{\pm}}$
their closures. If
$X \equiv \{x_{1},\cdots, x_{m}\} \in \R^{4m}$
and
$Y \equiv \{y_{1},\cdots, y_{n}\} \in \R^{4m}$
are such that
$
x_{i} - y_{j} \not\in \overline{V^{-}}, \quad \forall i=1,\dots,m,\quad
j=1,\dots,n
$
we use the notation
$X \geq Y.$
If
$
x_{i} - y_{j} \not\in \overline{V^{+}} \cup \overline{V^{-}}, 
\quad \forall i=1,\dots,m,\quad
j=1,\dots,n
$
we use the notations:
$X \sim Y.$
We use the compact notation
$T_{J}(X) \equiv T_{j_{1},\dots,j_{n}}(x_{1},\cdots, x_{n})$
with the convention
\be
T_{\emptyset}(\emptyset) \equiv {\bf 1}
\label{empty}
\ee
and by
$XY$
we mean the juxtaposition of the elements of $X$ and $Y$. Then the causality
axiom writes as follows:
\be
T_{J_{1}J_{2}}(X_{1}X_{2}) = T_{J_{1}}(X_{1}) T_{J_{2}}(X_{2}), 
\quad \forall X_{1} \geq X_{2};
\label{causality}
\ee
here 
$J_{i}$
are the indices corresponding to the the coordinates
$X_{i}$
i.e
$J_{i} \equiv pr(X_{i}), \quad i = 1, 2.$

>From (\ref{causality}) one can derive easily:
\be
[T_{J_{1}}(X_{1}), T_{J_{2}}(X_{2})] = 0, \quad {\rm if} 
\quad X_{1} \sim X_{2}.
\label{commute}
\ee

\item 
The {\it unitarity} of the $S$-matrix can be expressed if one introduces,
the formal series:
\be
\bar{S}({\bf g})=1 + \sum_{n=1}^\infty{(-i)^{n}\over n!}\int_{\R^{4n}}
dx_{1}\cdots dx_{n}\ \bar{T}_{j_{1},\dots,j_{n}}(x_{1},\cdots, x_{n}) 
g_{j_{1}}(x_{1})\cdots g_{j_{n}}(x_{n}),
\label{barS}
\ee
where, by definition:
\be
(-1)^{|X|} \bar{T}_{J}(X) \equiv \sum_{r=1}^{|X|} (-1)^{r} 
\sum_{X_{1},\dots,X_{r} \in Part(X)}
T_{J_{1}}(X_{1})\cdots T_{J_{r}}(X_{r});
\label{antichrono}
\ee 
here $X_{1},\cdots,X_{r}$ is a partition of $X$, $|X|$ is the cardinal of
the set $X$ and the sum runs over all partitions. In the lowest orders we have:
\be
\bar{T}_{j}(x) = T_{j}(x)
\ee
and
\be
\bar{T}_{j_{1}j_{2}}(x_{1},x_{2}) = 
- T_{j_{1}j_{2}}(x_{1},x_{2}) + T_{j_{1}}(x_{1}) T_{j_{2}}(x_{2}) + 
T_{j_{2}}(x_{2}) T_{j_{1}}(x_{1}).
\ee

One calls the operator-valued distributions
$\bar{T}_{j_{1},\dots,j_{n}}(x_{1},\dots,x_{n})$
{\it anti-chronological products}. The series (\ref{barS}) is the inverse of
the series (\ref{S}) i.e. we have:
\be
\bar{S}({\bf g}) = S({\bf g})^{-1}
\ee
in the sense of formal series. Then the unitarity axiom is then:
\be
\bar{T}_{J}(X) = T_{J}(X)^{\dagger}, \quad \forall X.
\label{unitarity}
\ee

One can show that the following relations are identically verified:
\be
\sum_{X_{1},X_{2} \in Part(X)}
(-1)^{|X_{1}|} T_{J_{1}}(X_{1}) \bar{T}_{J_{2}}(X_{2}) = 
\sum_{X_{1},X_{2} \in Part(X)} 
(-1)^{|X_{1}|} \bar{T}_{J_{1}}(X_{1}) T_{J_{2}}(X_{2}) = 0.
\label{unit}
\ee

Also one has, similarly to (\ref{causality}):
\be
\bar{T}_{J_{1}J_{2}}(X_{1}X_{2}) = 
\bar{T}_{J_{2}}(X_{2}) \bar{T}_{J_{1}}(X_{1}), 
\quad \forall X_{1} \geq X_{2}.
\label{bar-causality}
\ee
\end{itemize}

A {\it renormalization theory} is the possibility to construct such a
$S$-matrix starting from the first order terms:
$
T_{j}(x), \quad j = 1,\dots, P
$
which are linearly independent Wick polynomials called {\it interaction
Lagrangians} which should verify the following axioms:
\be
{\cal U}_{a,A} T_{j}(x) {\cal U}^{-1}_{a,A} = 
S(A^{-1})_{jk} \quad T_{k}(\delta(A)\cdot x+a),
\quad \forall A \in SL(2,\C), \quad \forall j = 1,\dots, P
\label{inv1}
\ee
\be
\left[T_{j}(x), T_{k}(y)\right] = 0, \quad \forall x,y \in \R^{4} \quad
s.t. \quad x \sim y, \quad \forall j, k = 1,\dots, P
\label{causality1}
\ee
and
\be
T_{j}(x)^{\dagger} = T_{j}(x), \quad \forall j = 1,\dots, P.
\label{unitarity1}
\ee

Usually, these requirements are supplemented by covariance with respect to some
discrete symmetries (like spatial and temporal inversions, or PCT), charge
conjugations or global invariance with respect to some Lie group of symmetry.
Some other restrictions follow from the requirement of the existence of the
adiabatic limit, at least in the weak sense.

The case of a {\it single Lagrangian} perturbation theory corresponds to 
$P = 1$.
In this case the expression
$T(x) = T_{1}(x)$
is the interaction Lagrangian and the chronological products are
$T(X) \equiv T_{1\dots 1}(X)$.

More generally, one can consider that the interaction Lagrangian is
\be
T(x) = \sum c_{j} T_{j}(x)
\label{one}
\ee
with 
$c_{j}$
some real constants. In this case, the chronological products of the theory are
\be
T(X) = \sum c_{j_{1}} \dots c_{j_{n}} T_{j_{1},\dots,j_{n}}(X).
\label{one-n}
\ee
\newpage

\subsection{Epstein-Glaser Induction}{\label{EG}}

We summarize the steps of the inductive construction of Epstein and Glaser
\cite{EG1}, \cite{Gr5}. Let the interaction Lagrangians 
$T_{j}(x), j = 1,\dots, P$
be some linearly independent Wick monomials acting in a certain Fock
space with
$\omega_{j}, \quad j = 1,\dots,P$
the corresponding canonical dimensions. We suppose that they generate
the space of all Wick monomials of canonical dimension less that $4$.
The causality property (\ref{causality1}) is fulfilled, but we must
make sure that we also have (\ref{inv1}) and (\ref{unitarity1}).

Moreover, a certain generalization of the preceding formalism is
needed in order express the operator-valued chronological product in
terms of numerical distributions \cite{EG1}. It is convenient let the
index $j$ to run from $0$ to $P$ and to give, by definition 
\be
T_{0} \equiv {\bf 1}.
\ee
Next, we define the summ
$j_{1} + j_{2}$
of two indices 
$j_{1}, j_{2} = 0,\dots, P$
through the relation
\be
T_{j_{1}+j_{2}}(x) = :T_{j_{1}}(x) T_{j_{2}}(x):
\ee
and then we extend the summation operation to $n$-uples of indices
$J = (j_{1},\dots,j_{n})$
componentwise.

We will use the notation
\be
\omega_{J} \equiv \sum_{j \in J} \omega_{j}
\ee
and we call it the {\it canonical dimension} of 
$T_{J}(X)$.

We suppose that we have constructed the chronological products 
$T_{j_{1},\dots,j_{p}}(x_{1},\cdots,x_{p})$
(for all
$p = 1, \dots, n - 1$)
having the following properties: (\ref{sym}), (\ref{causality}) and
(\ref{unitarity}) for 
$p \leq n - 1$,
(\ref{causality}) for
$|X_{1}| + |X_{2}| \leq n - 1$
and (\ref{commute}) for
$|X_{1}|, |X_{2}| \leq n - 1$.
Moreover, we suppose that we have the following {\it Wick expansion}
of the chronological products:
\be
T_{J}(X) = \sum_{K+L=J} t_{K}(X) :T_{l_{1}}(x_{1})\cdots T_{l_{n}}(x_{n}):
\label{wick-chrono}
\ee
for 
$|X| \leq n - 1$;
here
$t_{K}(X)$
are numerical distributions (called 
{\it renormalized Feynman amplitudes}) with degree of singularity restricted by
the following relation:
\be
\omega(t_{K}) \leq \omega_{K} - 4(n-1).
\label{deg-chrono}
\ee

Let us notice that from (\ref{wick-chrono}) we have:
\be
t_{J}(X) = \left( \Phi_{0}, T_{J}(X) \Phi_{0}\right).
\label{average-chrono}
\ee 

We want to construct the distribution-valued operators
$T_{J}(X), \quad |X| = n$
such that the the properties above go from $1$ to $n$.
Here are the steps of the construction.
\begin{enumerate}

\item
One constructs from
$T_{J}(X), \quad |X| \leq n - 1$
the expressions
$\bar{T}_{J}(X), \quad |X| \leq n - 1$
according to (\ref{antichrono}) and proves the properties
(\ref{bar-causality}) for
$|X_{1}| + |X_{2}| \leq n - 1$.

\item
Next, we defines the expressions:
\be
A_{j_{1},\dots,j_{n}}'(x_{1},\dots,x_{n-1};x_{n}) \equiv 
{\sum}'_{X_{1},X_{2} \in Part(X)}
(-1)^{|X_{2}|} T_{J_{1}}(X_{1}) \bar{T}_{J_{2}}(X_{2}),
\ee
\be
R_{j_{1},\dots,j_{n}}'(x_{1},\dots,x_{n-1};x_{n}) \equiv 
{\sum}'_{X_{1},X_{2} \in Part(X)} 
(-1)^{|X_{2}|} \bar{T}_{J_{1}}(X_{1}) T_{J_{2}}(X_{2}) 
\ee
where the sum
${\sum}'$
goes over the partitions of
$X = \{x_{1},\dots,x_{n}\}$
such that
$X_{2} \not= \emptyset, \quad x_{n} \in X_{1}.$

Next, we construct the expression
\be
D_{j_{1},\dots,j_{n}}(x_{1},\dots,x_{n-1};x_{n}) \equiv 
A_{j_{1},\dots,j_{n}}'(x_{1},\dots,x_{n-1};x_{n}) -
R_{j_{1},\dots,j_{n}}'(x_{1},\dots,x_{n-1};x_{n}).
\label{com-D}
\ee
and prove that it has causal support i.e.
$supp(D_{j_{1},\dots,j_{n}}(x_{1},\dots,x_{n-1};x_{n})) 
\subset \Gamma^{+}(x_{n}) \cup \Gamma^{-}(x_{n})$
where we use standard notations:
\be
\Gamma^{\pm}(x_{n}) \equiv \{ (x_{1},\dots,x_{n}) \in (\R^{4})^{n} |
x_{i} - x_{n} \in V^{\pm} , \quad \forall i = 1, \dots, n-1\}.
\ee
\item
The distributions
$D_{J}(X)$
can be written in a formula similar to (\ref{wick-chrono}):
\be
D_{J}(X) = \sum_{K+L=J} d_{K}(X) :T_{l_{1}}(x_{1})\cdots T_{l_{n}}(x_{n}):
\label{wick-d}
\ee
where
$d_{K}(X)$
are numerical distributions; in analogy to (\ref{average-chrono}) we have:
\be
d_{J}(X) = \left( \Phi_{0}, D_{J}(X) \Phi_{0}\right).
\label{average-d}
\ee 

It follows that the numerical distributions
$d_{J}(X)$
have causal support i.e
$
supp(d_{J}(X)) \subset \Gamma^{+}(x_{n}) \cup \Gamma^{-}(x_{n})
$
and are 
$SL(2,\C)$-invariant. Moreover, their degree of singularity is restricted by
\be
\omega(d_{K}) \leq \omega_{K} - 4(n-1).
\label{deg-d}
\ee

\item
There exists a causal splitting
\be
d = a - r, \quad supp(a) \subset \Gamma^{+}(x_{n}), \quad 
supp(r) \subset \Gamma^{-}(x_{n})
\ee
which is also
$SL(2,\C)$-invariant and such that the order of the singularity is preserved.
So, there exists a 
$SL(2,\C)$-covariant causal splitting:
\be
D_{j_{1},\dots,j_{n}}(x_{1},\dots,x_{n-1};x_{n}) =
A_{j_{1},\dots,j_{n}}(x_{1},\dots,x_{n-1};x_{n}) - 
R_{j_{1},\dots,j_{n}}(x_{1},\dots,x_{n-1};x_{n})
\label{decD}
\ee
with
$supp(A_{j_{1},\dots,j_{n}}(x_{1},\dots,x_{n-1};x_{n})) \subset 
\Gamma^{+}(x_{n})$
and
$supp(R_{j_{1},\dots,j_{n}}(x_{1},\dots,x_{n-1};x_{n})) \subset 
\Gamma^{-}(x_{n})$.

The expressions
$A_{n}$ 
and
$R_{n}$
are the {\it advanced} (resp. {\it retarded}) products.
\item
We have the relation
\be
D_{j_{1},\dots,j_{n}}(x_{1},\dots,x_{n-1};x_{n})^{\dagger} = (-1)^{n-1}
D_{j_{1},\dots,j_{n}}(x_{1},\dots,x_{n-1};x_{n}).
\ee
The causal splitting obtained above can be chosen such that
\be
A_{j_{1},\dots,j_{n}}(x_{1},\dots,x_{n-1};x_{n})^{\dagger} = (-1)^{n-1}
A_{j_{1},\dots,j_{n}}(x_{1},\dots,x_{n-1};x_{n}).
\ee

\item
Let us define
\bea
T_{j_{1},\dots,j_{n}}(x_{1},\cdots, x_{n}) \equiv 
A_{j_{1},\dots,j_{n}}(x_{1},\cdots, x_{n-1};x_{n}) - 
A_{j_{1},\dots,j_{n}}'(x_{1},\cdots, x_{n-1};x_{n}) 
\nonumber \\ \equiv 
R_{j_{1},\dots,j_{n}}(x_{1},\cdots, x_{n-1};x_{n}) - 
R_{j_{1},\dots,j_{n}}'(x_{1},\cdots, x_{n-1};x_{n}).
\label{chronos-n}
\eea
Then these expressions satisfy the 
$SL(2,\C)$-covariance, 
causality and unitarity conditions (\ref{invariance}) (\ref{causality}) 
(\ref{commute}) and (\ref{unitarity}) for
$p = n$.
If we substitute
\be
T_{j_{1},\dots,j_{n}}(x_{1},\cdots, x_{n}) \rightarrow {1 \over n!}
\sum_{\pi} T_{j_{\pi(1)},\dots,j_{\pi(n)}}(x_{\pi(1)},\cdots, x_{\pi(n)})
\ee
where the sum runs over all permutations of the numbers
$\{1, \dots, n\}$
then we also have the symmetry axiom (\ref{sym}). It is easy to see
that the induction hypothesis is verified by the operators
$T_{J}(X), \quad |X| = n$
constructed in this way.
\end{enumerate}
\newpage

\subsection{The Arbitrariness of the Chronological Products\label{arb}}

This problem was addressed in \cite{BS} and \cite{Pi}, as we have
already mention in the Introduction. We prefer to give an independent 
formulation based on the multi-Lagrangian Epstein-Glaser scheme
presented above. We consider two solutions of the Bogoliubov axioms
with the same ``initial conditions" 
$T_{j}, \quad j = 1,\dots, P$
chosen as a basis in the space of Wick monomials of degree
$4$.
We introduce the following notation: if 
$X = \{x_{1},\dots,x_{n}\}$ 
then
\be
\delta(X) \equiv \delta(x_{1}-x_{n}) \cdots \delta(x_{n-1}-x_{n}).
\ee

We note the identity:
\be
\sum_{x_{l} \in X} {\partial \over \partial x^{\mu}_{l}} \quad \delta(X) = 0.
\label{delta}
\ee

Now, we have the following result:
\begin{thm}
Let 
$T_{J}(X)$
and
$\tilde{T}_{J}(X)$
be two solutions of the Bogoliubov axioms such that
$T_{j}(x) = \tilde{T}_{j}(x), \quad \forall j = 1,\dots, P$
and both verify the restriction (\ref{deg-chrono}). Then we have the relation:
\be
\tilde{T}_{J}(X) = T_{J}(X) + \sum_{r=1}^{|X|-1} {1\over r!}
\sum_{X_{1},\dots,X_{r} \in Part(X)} 
P_{J_{1};k_{1}}(X_{1})\cdots P_{J_{r};k_{r}}(X_{r})
T_{k_{1},\dots,k_{r}}(x_{i_{1}},\dots,x_{i_{r}}), \quad \forall |X| \geq 2
\label{arbitrar}
\ee
where summation over the indices
$k_{1},\dots,k_{r} = 0,\dots, P$
is understood,
$P_{J;k}(X)$ 
are distributions of the form
\be
P_{J;k}(X) = p_{J;k}(\partial) \delta(X)
\label{p-delta}
\ee
with 
$p_{J;k}(\partial)$
a Lorentz covariant polynomial with constant coefficients in the partial
derivatives restricted by:
\be
deg(p_{J;k}) + \omega_{k} \leq \omega_{J} - 4(n-1)
\ee
and
$x_{i_{p}} \in X_{p}, \forall p = 1,\dots, r$.
In the preceding equation, the convention 
\be
P_{j;k}(X) \equiv \delta_{jk}, \quad |X| = 1
\label{conv}
\ee
is understood.
\label{arbitrar-thm}
\end{thm}

{\bf Proof:}
We use complete induction. For 
$n = 2$
one obtains a possible expression
$T_{j_{1}j_{2}}(x_{1},x_{2})$
by causally splitting the distribution
$D_{j_{1}j_{2}}(x_{1},x_{2}) = [T_{j_{1}}(x_{1}),T_{j_{2}}(x_{2})]$.
According to a general result in distribution splitting theory, two such
splitting differ by a distribution with support in the set
$\{x_{1} = x_{2}\}$
of the type 
$
\sum_{k=0}^{P} [p_{j_{1}j_{2};k}(\partial) \delta(x_{1}-x_{2})] T_{k}(x_{2});
$
the limitation
$
deg(p_{j_{1}j_{2};k}) + \omega_{k} \leq 
\omega_{j_{1}} + \omega_{j_{2}} - 4
$
follows from the restrictions (\ref{deg-chrono}). The Lorentz
covariance follows if we make the distribution splitting in a
covariant way, 
which is known to be possible. 

We suppose that we have the expressions
$P_{J;k}(X), \quad |X| \leq n-1$
such that the formula from the statement is valid for 
$|X| \leq n-1$;
we prove the formula for 
$|X| = n$.
It is convenient to observe that one can write the formula
(\ref{arbitrar}) as follows:
\be
\tilde{T}_{J}(X) = \sum_{r=1}^{|X|} {1\over r!}
\sum_{X_{1},\dots,X_{r} \in Part(X)} 
P_{J_{1};k_{1}}(X_{1})\cdots P_{J_{r};k_{r}}(X_{r})
T_{k_{1},\dots,k_{r}}(x_{i_{1}},\dots,x_{i_{r}}), \quad \forall |X| \geq 2
\label{arbitrar1}
\ee
where the convention (\ref{conv}) has been used. So, by the induction 
hypothesis, we have the previous relation for 
$|X| \leq n - 1$.

Let us consider in this case the expression
\be
\Delta_{J}(X) \equiv \tilde{T}_{J}(X) - \sum_{r=2}^{|X|} 
{1\over r!} \sum_{X_{1},\dots,X_{r} \in Part(X)} 
P_{J_{1};k_{1}}(X_{1})\cdots P_{J_{r};k_{r}}(X_{r})
T_{k_{1},\dots,k_{r}}(x_{i_{1}},\dots,x_{i_{r}})
\ee
and show that it has the support in the set
$x_{1} = x_{2} = \cdots = x_{n}$.
For this, let us suppose that the point
$(x_{1},\dots,x_{n})$
is outside this set. Then one can find a Cauchy surface separating this set in
two non-void subsets $Y$ and $Z$ such that
$[Y] \geq [Z]$.
Because of the symmetry axiom (\ref{sym})  we can suppose, without loosing 
generality, that
$Y = \{x_{1},\dots,x_{i}\}$
and
$Z = \{x_{i+1},\dots,x_{n}\}$.
In that case, let us notice that in the sum appearing in the preceding formula
we can have non-zero contributions only from those partitions
$X_{1},\dots,X_{r}$
such that for every
$p = 1,\dots, r$
we have either
$X_{p} \subset Y$
or
$X_{p} \subset Z$.
This means that for such a choice of
$(x_{1},\dots,x_{n})$
we have:
\bea
\Delta_{J}(X) \equiv \tilde{T}_{J}(X) -
\sum_{2 \leq s+t\leq |X|} 
\sum_{Y_{1},\dots,Y_{s} \in Part(Y)} {1\over s!}
\sum_{Z_{1},\dots,Z_{t} \in Part(Z)} {1\over t!}
\nonumber \\
P_{J_{1};k_{1}}(Y_{1})\cdots P_{J_{s};k_{s}}(Y_{s})
P_{J_{s+1};k_{s+1}}(Z_{1})\cdots P_{J_{s+t};k_{s+t}}(Z_{t})
T_{k_{1},\dots,k_{s+t}}(x_{i_{1}},\dots,x_{i_{s+t}})
\eea
with
$x_{i_{p}} \in Y_{p}, \forall p = 1,\dots, s$
and
$x_{i_{s+p}} \in Z_{p}, \forall p = 1,\dots, t$.
Now, we can use in the right hand side the causality property (\ref{causality})
for the chronological products
$
T_{J}(X)
$
and
$\tilde{T}_{J}(X)$..
We have easily get
$\Delta_{J}(X) = 0$.
The support property of the distribution
$\Delta_{J}(X)$
is proved. Using Wick theorem and well known facts about the structure of 
numerical distribution with support included in the set
$x_{1} = x_{2} = \cdots = x_{n}$
we get the formula (\ref{arbitrar}) for
$|X| = n$.
The Lorentz covariance follows like in the case 
$n = 2$.
This finished the induction.
$\qed$

It is clear now why do we need the multi-Lagrangian generalisation of
Epstein-Glaser formalism. Even if we work in a theory with a single
Lagrangian, the best we can do is to choose it among the set of linearly
independent Wick polynomials
$T_{j}$
say,
$T(x) = T_{1}(x)$
and the usual chronological products of a single Lagrangian theory are
$T(X) = T_{1\dots 1}(X)$
(see the end of Subsection \ref{bogoliubov}). To sets of chronological products
$T(X)$
and
$\tilde{T}(X)$
with the same ``initial condition"
$T(x) = \tilde{T}(x)$
will be connected by a formula of the following type:
\be
\tilde{T}(X) = T(X) + \sum_{r=1}^{|X|-1} {1\over r!}
\sum_{X_{1},\dots,X_{r} \in Part(X)} 
P_{k_{1}}(X_{1})\cdots P_{k_{r}}(X_{r})
T_{k_{1},\dots,k_{r}}(x_{i_{1}},\dots,x_{i_{r}}), \quad \forall |X| \geq 2
\label{tt1}
\ee
where we have denoted
$P_{k}(X) \equiv P_{\{1\dots 1\};k}(X)$
with $|X|$ entries of the figure $1$.
So, in the difference between two solutions of the problem will certainly
appear other chronological products that
$T(X)$.
\newpage

\section{Dilation Covariance of the Chronological Products\label{scale-chrono}}

\subsection{A General Characterization of Dilation Properties}

We will use the result from the preceding Subsection to study the generic
behaviour of the chronological products with respect to the dilation invariance
operators which was defined in Section 2. More explicitly, we consider
a certain choice for the chronological products and we will emphasize the mass 
dependence in an obvious way:
$T_{J}(X;m)$;
these expressions are not completely fixed for 
$|X| \geq 2$
because there is the possibility of finite renormalizations. Nevertheless, 
we make a concrete choice in accordance with Bogoliubov axioms and we have:
\begin{prop}
We suppose that the framework from the preceding Section is valid. Then the
following relations are valid for all
$|X| \geq 2$:
\bea
{\cal U}_{\lambda}T_{J}(X;m) {\cal U}_{\lambda}^{-1} =
\lambda^{\omega_{J}} T_{J}(\lambda X;\lambda^{-1}m) 
\nonumber \\
+ \sum_{r=1}^{|X|-1} {1\over r!} \sum_{X_{1},\dots,X_{r} \in Part(X)} 
P_{J_{1};k_{1};m;\lambda}(\lambda X_{1})\cdots 
P_{J_{r};k_{r};m;\lambda}(\lambda X_{r})
T_{k_{1},\dots,k_{r}}(\lambda x_{i_{1}},\dots,\lambda x_{i_{r}};\lambda^{-1} m)
\label{scale-n}
\eea
where the distributions
$P_{J;k;m;\lambda}(X)$
are of the form
\be
P_{J;k;m;\lambda}(X) = \sum_{\alpha} c_{J;k;\alpha}(\lambda,m) 
\partial^{\alpha} \delta(X);
\label{polyn-lambda}
\ee
here
$\alpha$
are multi-indices and
$|\alpha|$
is the corresponding length. Moreover, the following relation is verified:
\be
P_{J;k;m;1} = 0 \quad \Longleftrightarrow c_{J;k;\alpha}(1,m) = 0.
\label{ini}
\ee
\label{scale-Tn}
\end{prop}

{\bf Proof:}
Let us consider the following expressions
\be
T^{\lambda}_{J}(X) \equiv \lambda^{\omega_{J}} T_{J}(X;\lambda^{-1} m), \quad
\tilde{T}^{\lambda}_{J}(X) \equiv 
{\cal U}_{\lambda}T_{J}(\lambda^{-1} X;m) {\cal U}_{\lambda}^{-1}, 
\quad \forall |X| \geq 2
\ee
both acting in the same Fock space:
${\cal F}_{\lambda^{-1}m}$
and having the same ``initial conditions"
\be
T^{\lambda}_{j}(x) \equiv \lambda^{\omega_{j}} T_{j}(x), \quad j = 1,\dots, P
\ee
due to (\ref{scale}).

Also, these expressions verify the Bogoliubov axioms: the unitarity and the 
causality are obvious, but for the Poincar\'e covariance one had to use the 
relation (\ref{reps}). We can apply theorem \ref{arbitrar-thm} and obtain that
the difference between the two expressions 
$\tilde{T}^{\lambda}_{J}(X)$ 
and
$T^{\lambda}_{J}(X)$ 
is a sum of the type appearing in the right hand side of the relation
(\ref{arbitrar}) but with the polynomials depending on the parameter 
$\lambda$.
If we make the substitution 
$X \rightarrow \lambda X$ 
we get the relation from the statement. If we take 
$\lambda = 1$
then we get (\ref{ini}).
$\qed$

\begin{rem}
If we change the choice of the chronological products
$T_{J}(X;m), \quad |X| \geq 2$
adding some finite renormalizations, then the distributions
$P_{J;k;m;\lambda}(X)$
will change also by some ``coboundary" contribution. We will use this
freedom later to simplify their form.
\end{rem}

The central result of this paper describes the explicit $\lambda$-dependence of
the polynomials appearing in the preceding proposition. We  study
separately the cases 
$m \not = 0$
and
$m = 0$.

\subsection{The Case $m \not= 0$\label{not}}

In this case the following result hold:

\begin{thm}
In the case
$m \not= 0$
one can choose the chronological products such that the following
relations are valid for any
$|X| \geq 2$:
\be
{\cal U}_{\lambda}\quad  T_{J}(X;m)\quad {\cal U}_{\lambda}^{-1} =
\lambda^{\omega_{J}}\quad T_{J}(\lambda X;\lambda^{-1} m).
\ee
\label{m-non-zero}
\end{thm}

{\bf Proof:}
Is done by induction.

(i) First, we consider the case
$|X| = 2$. We start from the relation (\ref{scale-n}) from the preceding
proposition and apply
${\cal U}_{\lambda'} \cdots {\cal U}_{\lambda'}^{-1}$.
We easily obtain the cocycle identity
\be
P_{J;k;m;\lambda\lambda'}(X) = 
\lambda^{\omega_{J}} P_{J;k;\lambda^{-1}m;\lambda'}(X) +
(\lambda')^{\omega_{k}} P_{J;k;m;\lambda}(\lambda'^{-1}X).
\label{co-2}
\ee

If we substitute here the expression (\ref{polyn-lambda}) for 
$P_{J;k;\lambda}(X)$
one finds out immediately from the preceding cocycle identity that we have
\be
c(\lambda\lambda';m) = \lambda^{\omega_{J}} c(\lambda';\lambda^{-1}m) +
(\lambda')^{4+\omega_{k}+|\alpha|} c(\lambda;m).
\label{c2}
\ee
where we omit for simplicity the dependence on
$J, k$
and
$\alpha$. 
More conveniently, one defines
\be
d(\lambda,m) \equiv \lambda^{-\omega_{J}} c(\lambda,m)
\label{cd}
\ee
and has the cocycle identity:
\be
d(\lambda\lambda',m) = d(\lambda',\lambda^{-1}m) + 
(\lambda')^{s} d(\lambda,m)
\label{cocycle-d}
\ee
where we have denoted
\be
s \equiv 4 + \omega_{k} + |\alpha| -\omega_{J}.
\label{s2}
\ee

>From (\ref{ini}) we have the ``initial condition":
\be
c(1,m) = 0 \quad \Longleftrightarrow \quad d(1,m) = 0.
\ee

The equation (\ref{cocycle-d}) can be analysed elementary: if we take
$m = \lambda$
the following  equation emerges:
\be
d(\lambda\lambda',\lambda) = d(\lambda',1) + 
(\lambda')^{s} d(\lambda,\lambda)
\ee
or, if we make
$\lambda = M$
and
$\lambda \lambda' = \Lambda$
we have
\be
d(\Lambda,M) = d\left({\Lambda\over M},1\right) + 
\left({\Lambda\over M}\right)^{s} d(M,M).
\label{lambda-M}
\ee

Now, we substitute this expression into the initial relation
(\ref{cocycle-d}) we immediately get
\be
d\left({\lambda\over m},1\right) = - 
\left({\lambda\over m}\right)^{s} d(\lambda^{-1}m,\lambda^{-1}m)
\ee
which makes sense because 
$m \not= 0$.
Finally, we substitute this expression into (\ref{lambda-M}) and
obtain the most general solution of the cohomological equation 
(\ref{cocycle-d})
\be
d(\lambda,m) =  \lambda^{s} p(m) - p(\lambda^{-1}m)
\ee
where
$p(m) = {1\over m^{s}} d(m,m)$.
This proves, in particular, that the general solution of the
cohomological equation (\ref{cocycle-d}) is in this case a coboundary. In the
end we have
\be
c_{J;k;\alpha}(\lambda,m) = 
\lambda^{4+|\alpha|+\omega_{k}} p_{J;k;\alpha}(m)
- \lambda^{\omega_{J}} p_{J;k;\alpha}(\lambda^{-1}m) 
\ee 
where
$p_{J;k;\alpha}$
are arbitrary functions on $m$. It follows that the polynomial 
$P_{J;k;m;\lambda}$
appearing in the relation (\ref{scale-n}) for 
$|X| = 2$ 
has the generic form
\be
P_{J;k;m\lambda}(X) = \sum_{\alpha}
\left[ \lambda^{4+|\alpha|+\omega_{k}} p_{J;k;\alpha}(m)
- \lambda^{\omega_{J}} p_{J;k;\alpha}(\lambda^{-1}m)\right]  
\partial^{\alpha} \delta(X).
\label{polyn-triv}
\ee

Now, it is an easy exercise to prove that this expression is a
``coboundary" in the sense that if we redefine the chronological
products
$T_{J}(X;m)$
in order $2$ according to
\be
\tilde{T}_{J}(X,m) \equiv T_{J}(X,m) - \sum_{\alpha} p_{J;k;\alpha}(m)
\left[ \partial^{\alpha} \delta(X)\right] T_{k}(x_{2},m) 
\ee
then we will have for
$|X| = 2$
a relation of the type (\ref{scale-n}) with
$P_{J;k;m;\lambda} \longrightarrow \tilde{P}_{J;k;m;\lambda} = 0$. 
This proves the assertion from the statement in the case
$|X| = 2$
that is, we can redefine the chronological products in the second
order such that we have the relation from the satement for 
$|X| = 2$.

(ii) We suppose that the formula from the statement is valid for 
$2 \leq |X| \leq n-1$
and we prove it for
$|X| = n$.
The induction hypothesis amounts to
\be
P_{J;k;m;\lambda}(X) = 0, \quad |X| \leq n - 1.
\ee

Then, we get from (\ref{scale-n}) for
$|X| = n$
the following relation
\be
{\cal U}_{\lambda}T_{J}(X;m) {\cal U}_{\lambda}^{-1} =
\lambda^{\omega_{J}} T_{J}(\lambda X;\lambda^{-1}m) 
+  P_{J;k;m;\lambda}(\lambda X) T_{k}(\lambda x_{n};\lambda^{-1}m)
\quad |X| = n
\ee
which is a relation of the same type as the relation for
$T_{J}(X,m), \quad |X| = 2$
obtained above. One can obtain a cohomological relation for the
polynomials 
$P_{J;k;m;\lambda}$
which coincides in fact with (\ref{co-2}). For the coefficients
$c_{J;k;\alpha}(\lambda,m)$
we will obtain again an equation of the type (\ref{c2}):
\be
c(\lambda\lambda';m) = \lambda^{\omega_{J}} c(\lambda';\lambda^{-1}m) +
(\lambda')^{4(|X|-1)+\omega_{k}+|\alpha|} c(\lambda;m).
\label{cn}
\ee

Then we find out that the relation (\ref{cocycle-d}) is valid with
(\ref{s2}) modified to
\be
s = 4(|X|-1) + |\alpha| + \omega_{k} -\omega_{J}.
\label{sn}
\ee 

In the end, the most general expression of the polynomial
$P_{J;k;m;\lambda}$
is
\be
P_{J;k;m\lambda}(X) = \sum_{\alpha}
\left[ \lambda^{4(|X|-1)+|\alpha|+\omega_{k}} p_{J;k;\alpha}(m)
- \lambda^{\omega_{J}} p_{J;k;\alpha}(\lambda^{-1}m)\right]  
\partial^{\alpha} \delta(X).
\label{polyn-triv-n}
\ee

As before, one can make
$P_{J;k;\lambda}(X) = 0$
by a suitable redefinition of the chronological products
$T_{J}(X,m), \quad |X| = n$.
The induction is finished.
$\qed$

\subsection{The Case $m = 0$\label{zero}}

We remark that in the relation
(\ref{scale-n}) the distributions 
$P_{J;k;0;\lambda}$
appear and for simplicity we can skip the entry $0$ from the index. We
adopt the same convetion for the coefficients
$c_{J;k;\alpha}(\lambda,0)$
appearing in the generic expression (\ref{polyn-lambda}).

In this case 
$m = 0$
we have a much more interesting result:
\begin{thm}
In the case
$m = 0$
one can redefine the chronological products in such a way that the
distributions 
$P_{J;k;\lambda}(X)$
are of the following form:
\be
P_{J;k;\lambda}(X) = \lambda^{\omega_{J}} ln(\lambda)
\sum_{|\alpha| = \omega_{J} - 4(|X|-1) -\omega_{k}}
c_{J;k;\alpha;\lambda}\partial^{\alpha} \delta(X). 
\ee
\label{log}
\end{thm}

{\bf Proof:}
As before, is done by induction.

(i) First, we consider the case
$|X| = 2$. 
We start from the relation (\ref{scale-n}) from the proposition
\ref{scale-Tn} and apply
${\cal U}_{\lambda'} \cdots {\cal U}_{\lambda'}^{-1}$.
We obtain the cocycle identity of the same type as (\ref{co-2}):
\be
P_{J;k;\lambda\lambda'}(X) = \lambda^{\omega_{J}} P_{J;k;\lambda'}(X) +
(\lambda')^{\omega_{k}} P_{J;k;\lambda}(\lambda'^{-1}X)
\label{co-2-0}
\ee
where now we have no mass dependence. Insetead of the relation
(\ref{c2}) we get: 
\be
c(\lambda\lambda') = \lambda^{\omega_{J}} c(\lambda') +
(\lambda')^{4+|\alpha|+\omega_{k}} c(\lambda).
\label{c2-0}
\ee

As before, one defines
$
d(\lambda)
$
according to (\ref{cd}) and has the cocycle identity:
\be
d(\lambda\lambda') = d(\lambda') + (\lambda')^{s} d(\lambda), \quad
s \equiv 4 + \omega_{k} + |\alpha| -\omega_{J}.
\label{cocycle-d-0}
\ee

Again we have from (\ref{ini}) the ``initial condition":
\be
c(1) = 0 \quad \Longleftrightarrow \quad d(1) = 0.
\ee

The equation (\ref{cocycle-d-0}) can be analysed elementary if we differentiate
with respect to 
$\lambda'$
and put
$\lambda' = 1.$
The following differential equation emerges:
\be
\lambda d'(\lambda) = d_{0} + s d(\lambda)
\label{dif-d}
\ee
where 
$
d_{0} \equiv d'(1).
$
We have two cases:

(a) \underline{$s \not= 0$}

The homogeneous equation
$
\lambda D'(\lambda) = s D(\lambda)
$
has the solution
$D(\lambda) = A \lambda^{s}$.
With the methods of variation of constants, we look for a solution of the
preceding equation of the form
$
d(\lambda) = A(\lambda) \lambda^{s}
$
with the initial condition
$A(1) = 0.$
The function
$A(\lambda)$
will verify the equation:
\be
A' = d_{0} \lambda^{-s-1}
\ee
with the solution
\be
A(\lambda) = {d_{0}\over s} (1 - \lambda^{-s})
\ee

>From here we get the solution
\be
d(\lambda) = {d_{0}\over s} (\lambda^{s} - 1)
\label{ddd}
\ee
which verifies identically the initial equation (\ref{cocycle-d-0}).

(b) \underline{ $s = 0$}

The equation (\ref{dif-d}) becomes:
\be
\lambda d'(\lambda) = d_{0}
\ee
with the solution
$
d(\lambda) = d_{0} \ln(\lambda)
$
which, again, identically verifies the initial equation (\ref{cocycle-d-0}). 
We get in this case the solution:
\be
c(\lambda) = d_{0} \lambda^{\omega_{J}} ln(\lambda).
\ee

In the end, we get, instead of the formula (\ref{polyn-triv}) 
\bea
P_{J;k;m\lambda}(X) = \sum_{|\alpha| \not= \omega_{J}-4-\omega_{k}}
\left[ \lambda^{4+|\alpha|+\omega_{k}} c_{J;k;\alpha}
- \lambda^{\omega_{J}} c_{J;k;\alpha}\right]  
\partial^{\alpha} \delta(X)
\nonumber \\
+ \sum_{|\alpha| = \omega_{J} - 4 -\omega_{k}} 
\lambda^{\omega_{J}} ln(\lambda) c_{J;k;\alpha}\partial^{\alpha} \delta(X).
\label{polyn-triv-0}
\eea

If we make a redefinition of the chronological products
$T_{J}(X,0), \quad |X| = 2$
we can get rid of the first sum in the preceding expression as in the
case 
$m \not= 0$. 
This means that one can take
\be
P_{J;k;m\lambda}(X) = \lambda^{\omega_{J}} ln(\lambda) 
\sum_{\alpha = \omega_{J} - 4 -\omega_{k}} 
c_{J;k;\alpha}\partial^{\alpha} \delta(X)
\label{polyn-triv-00}
\ee
which proves the assertion from the statement in the case
$|X| = 2.$

(ii) We suppose that the formula from the statement is valid for 
$2 \leq |X| \leq n-1$
and we prove it for
$|X| = n.$
We establish a cocycle identity for 
$P_{J;k;\lambda}(X),\quad |X| = n.$
Instead of (\ref{co-2-0}) we obtain in the same way:
\bea
P_{J;k;\lambda\lambda'}(X) = \lambda^{\omega_{J}} P_{J;k;\lambda'}(X) +
(\lambda')^{\omega_{k}} P_{J;k;\lambda}(\lambda'^{-1}X) 
\nonumber \\
+ \sum_{r=2}^{|X|-1} {1\over r!} \sum_{X_{1},\dots,X_{r} \in Part(X)} 
P_{J_{1};m_{1};\lambda}(\lambda'^{-1} X_{1})\cdots 
P_{J_{r};m_{r};\lambda}(\lambda'^{-1} X_{r})
P_{m_{1},\dots,m_{r};k;\lambda'}(x_{i_{1}},\dots, x_{i_{r}})
\label{co-n}
\eea

This relation goes into (\ref{co-2}) for 
$n = 2$
because the sum disappears. The preceding relation gives, instead of
(\ref{c2-0}) the following:
\be
c(\lambda\lambda') = \lambda^{\omega_{J}} c(\lambda') +
(\lambda')^{4(|X|-1)+\omega_{k}+|\alpha|} c(\lambda)
+ (\lambda\lambda')^{\omega_{J}} ln(\lambda')
\sum_{r=2}^{|X|-1} c_{r} ln^{r}(\lambda)
\ee
where, again, the multi-index 
$\alpha$
was omitted and
$c_{r}$
are some constants; their value will not be needed. If we define the function
$
d(\lambda) 
$
by (\ref{cd}) we get: 
\bea
d(\lambda\lambda') = d(\lambda') + (\lambda')^{s} d(\lambda)
+ ln(\lambda') \sum_{r=2}^{|X|-1} c_{r} ln^{r}(\lambda),
\nonumber \\
s \equiv 4(|X|-1) + \omega_{k} + |\alpha| -\omega_{J}.
\label{cocycle-d-n-0}
\eea

We know certainly from the general theorem \ref{scale-Tn} that this
equations must have solutions. The only problem is to determine the 
$\lambda$ dependence from the preceding equation.
As before we get from this relation the differential equation:
\be
\lambda d'(\lambda) = d_{0} + s d(\lambda)
+ \sum_{r=2}^{|X|-1} c_{r} ln^{r}(\lambda)
\label{dif-d-n}
\ee

We have the same cases as before.

(a) \underline{$s \not= 0$}

The homogeneous equation is again
$
\lambda D'(\lambda) = s D(\lambda)
$
with the  the solution
$D(\lambda) = A \lambda^{s}$.
If we look for a solution of the equation (\ref{dif-d-n}) of the form
$
d(\lambda) = A(\lambda) \lambda^{s}
$
with the initial condition
$A(1) = 0$
we get for
$A(\lambda)$
the equation:
\be
A' =  \lambda^{-s-1} 
\left[ d_{0} + \sum_{r=2}^{|X|-1} c_{r} ln^{r}(\lambda)\right].
\ee

In the end, we get, instead of (\ref{ddd}) the following expression for
the functions $d$:
\be
d(\lambda) = {d_{0}\over s} (\lambda^{s} - 1) 
+ \sum_{r=1}^{|X|-1} a_{r} ln^{r}(\lambda)
\ee
with 
$
a_{r}
$
some constants. We substitute in the original equation
(\ref{cocycle-d-n-0}) for the function $d$ and obtain that the only
possibility is to have 
$
a_{r} = 0
$
(and 
$
c_{r} = 0$)
so we have in fact the solution:
\be
d(\lambda) = {d_{0}\over s} (\lambda^{s} - 1). 
\ee

(b) \underline{$s = 0$}

The equation (\ref{dif-d}) becomes:
\be
\lambda d'(\lambda) = d_{0} + \sum_{r=2}^{|X|-1} c_{r} ln^{r}(\lambda).
\ee
with the solution
\be
d(\lambda) = d_{0} \ln(\lambda) + 
\sum_{r=2}^{|X|-1} {c_{r}\over r+1} ln^{r+1}(\lambda).
\ee

We substitute in the initial equation (\ref{cocycle-d-n-0}) and obtain that 
$
c_{r} = 0
$
so
\be
d(\lambda) = d_{0} \ln(\lambda) 
\quad \Longleftrightarrow \quad
c(\lambda) = d_{0} \lambda^{\omega_{J}} \ln(\lambda).
\ee

It follows that the most general expression of the polynomials 
$P_{J;k;\lambda}(X),\quad |X| = n$
is
\bea
P_{J;k;m\lambda}(X) = \sum_{|\alpha|\not= \omega_{J}-4(|X|-1)-\omega_{k}}
\left[ \lambda^{4(|X|-1)+|\alpha|+\omega_{k}} c_{J;k;\alpha}
- \lambda^{\omega_{J}} c_{J;k;\alpha}\right]  
\partial^{\alpha} \delta(X)
\nonumber \\
+ \sum_{|\alpha| = \omega_{J} - 4(|X|-1) -\omega_{k}} 
\lambda^{\omega_{J}} ln(\lambda) c_{J;k;\alpha}\partial^{\alpha} \delta(X).
\label{polyn-triv-0-n}
\eea

If we make a redefinition of the chronological products
$T_{J}(X,0), \quad |X| = n$
we can get rid of the first sum in the preceding expression as in the
case 
$m \not= 0$. 
This means that one can take
\be
P_{J;k;m\lambda}(X) = 
\lambda^{\omega_{J}} ln(\lambda) 
\sum_{|\alpha| = \omega_{J} - 4(|X|-1) -\omega_{k}} 
c_{J;k;\alpha}\partial^{\alpha} \delta(X)
\label{polyn-triv-0n}
\ee
which proves the assertion from the statement in the case
$|X| = n.$
The induction is finished.
$\qed$

If we substitute the preceding result into the proposition \ref{scale-Tn} we 
get the following result:
\begin{thm}
The following relations are valid for any
$|X| \geq 2$:
\bea
{\cal U}_{\lambda}T_{J}(X;0) {\cal U}_{\lambda}^{-1} =
\lambda^{\omega_{J}} [T_{J}(\lambda X;0) +
\nonumber \\
\sum_{r=1}^{|X|-1} {ln^{r}(\lambda)\over r!} 
\sum_{X_{1},\dots,X_{r} \in Part(X)} 
\lambda^{-(\omega_{k_{1}} + \cdots +\omega_{k_{r}})}
P_{J_{1};k_{1}}(X_{1})\cdots P_{J_{r};k_{r}}(X_{r}) \times
\nonumber \\
T_{k_{1},\dots,k_{r}}
(\lambda x_{i_{1}},\dots,\lambda x_{i_{r}};0)]
\label{scale-ln}
\eea
where the distributions
$P_{J;k}(X)$
are of the form
\be
P_{J;k}(X) = \sum_{|\alpha| = \omega_{J} -4(|X|-1) -\omega_{k}}
c_{J;k;\alpha}\partial^{\alpha} \delta(X).
\ee

We also have:
\bea
{\cal U}_{\lambda}\bar{T}_{J}(X;0) {\cal U}_{\lambda}^{-1} =
\lambda^{\omega_{J}} [\bar{T}_{J}(\lambda X;0) +
\nonumber \\
\sum_{r=1}^{|X|-1} {ln^{r}(\lambda)\over r!} 
\sum_{X_{1},\dots,X_{r} \in Part(X)} 
\lambda^{-(\omega_{k_{1}} + \cdots + \omega_{k_{r}})}
\bar{P}_{J_{1};k_{1}}(X_{1})\cdots \bar{P}_{J_{r};k_{r}}(X_{r}) \times
\nonumber \\
\bar{T}_{k_{1},\dots,k_{r}}
(\lambda x_{i_{1}},\dots,\lambda x_{i_{r}};0)].
\eea
\label{logarithm}
\end{thm}

Let us also remark that in the case 
$m = 0$
the operators
${\cal U}_{\lambda}$
act in the same Fock space
${\cal F}^{+}_{0}$
and so, they form a unitary representation of the dilation group
$\R_{+}$.
This means that we can define the infinitesimal generatios of the 
dilations: let us consider the continuous unitary representation of
the additive group $\R$ given by
\be
V_{\chi} \equiv {\cal U}_{exp(\chi)}
\ee
and denote by $D$ its infinitesimal generator obtained via Stone-von-Neumann
theorem:
\be
V_{\chi} = e^{i \chi D}.
\ee

Then we have from (\ref{scale-phi}) the following commutation relation:
\be
\left[ D, \phi(x) \right] \sim -i 
\left( 1 + x^{\mu} {\partial \over \partial x^{\mu}} \right) \phi(x).
\label{inf}
\ee

The infinitesimal form of the relations (\ref{scale}) and
(\ref{scale-ln}) are:
\be
\left[ D, W(x) \right] \sim -i 
\left[ \omega(W) + x^{\mu} {\partial \over \partial x^{\mu}} \right] W(x).
\label{DW}
\ee
and  respectively:
\be
[D, T_{J}(X)] \sim -i \left(\omega_{J} + \sum_{l \in X} x^{\mu} 
{\partial \over \partial x_{l}^{\mu}}\right) T_{J}(X) -i
\sum P_{J;k}(X) T_{k}(x_{n}).
\label{scale-ln-infi}
\ee

%\newpage
\subsection{Scaling Properties of the Renormalized Feynman Amplitudes
\label{feynman}}

We translate the preceding results for the renormalized Feynman
amplitudes. From this analysis one can obtain the asymptotic behaviour
of these amplitudes as it is done in the classic paper of Weinberg \cite{We}. 
We use the expression (\ref{wick-chrono}) for 
$T_{J}(X;m)$ 
emphasizing the mass dependence: 
\be
T_{J}(X;m) = \sum_{K+L=J} t_{K}(X;m)
:T_{l_{1}}(x_{1};m) \cdots T_{l_{n}}(x_{n};m):
\label{t-wick}
\ee
and we have:
\begin{thm}
The following relations are verified:

(1) in the case 
$m \not= 0$:
\be
t_{J}(X;m) = \lambda^{\omega_{J}} 
t_{J}(\lambda X;\lambda^{-1} m);
\ee

(2) in the case 
$m = 0$:
\bea
t_{J}(X;0) = \lambda^{\omega_{J}} 
[t_{J}(\lambda X;0) +
\nonumber \\
\sum_{r=1}^{|X|-1} {ln^{r}(\lambda)\over r!} 
\sum_{X_{1},\dots,X_{r} \in Part(X)} 
\lambda^{-(\omega_{k_{1}} + \cdots +\omega_{k_{r}})}
P_{J_{1};k_{1}}(X_{1})\cdots P_{J_{r};k_{r}}(X_{r})\times
\nonumber \\
t_{k_{1},\dots,k_{r}}
(\lambda x_{i_{1}},\dots,\lambda x_{i_{r}};0)].
\eea
\label{scale-ln-feynman}
\end{thm}

The proof is done using the formula (\ref{average-chrono}) into the
relations (\ref{m-non-zero}) and (\ref{scale-ln}). The preceding
theorem elucidates the logarithmic
behaviour of the renormalized Feynman amplitudes in the case
$m = 0$.
Presumably, the terms proportional with 
$ln^{r}$ 
correspond to graphs with $r$ loops. 

One can obtain the infinitesimal form of the preceding relation: we make 
$\lambda = e^{\chi}$,
differentiate with respect to the variable $\chi$ and put
$\chi = 0$.
If we take into account that
\be
t_{j}(x) = \delta_{j,0}
\ee
the following relations emerges:

(1) for
$m \not= 0$:
\be
\left( \sum_{l=1}^{n} x^{\mu}_{l} {\partial \over \partial x^{\mu}_{l}} -
m {\partial \over \partial m} + \omega_{J} \right) t_{J;K}(X;m) = 0.
\ee

If we take into account translation invariance, we can express the Feynman
amplitudes 
$t_{J}(X;m)$
in terms of translation-invariant variables:
$
\xi_{i} \equiv x_{i} - x_{n}, \quad i = 1,\dots, n-1
$
and we have:
\be
\left( \sum_{l=1}^{n-1} \xi^{\mu}_{l} {\partial \over \partial \xi^{\mu}_{l}} -
m {\partial \over \partial m} + \omega_{J} \right) t_{J}(\Xi;m) = 0
\label{CS-xi-m}
\ee
or, for the Fourier transform:
\be
\left( \sum_{l=1}^{n-1} p^{\mu}_{l} {\partial \over \partial p^{\mu}_{l}} +
m {\partial \over \partial m} - \omega_{J} \right) 
\tilde{t}_{J;K}(P;m) = 0.
\label{CS-p-m}
\ee

(2) for
$ m =0$:
\be
\left( \sum_{l=1}^{n} x^{\mu}_{l} {\partial \over \partial x^{\mu}_{l}} 
+ \omega_{J} \right) t_{J;K}(X;m) + P_{J;0}(X) = 0
\ee
or, in translationally invariant variables:
\be
\left( \sum_{l=1}^{n-1} \xi^{\mu}_{l} {\partial \over \partial \xi^{\mu}_{l}} 
+ \omega_{J} \right) t_{J}(\Xi;m) + P_{J;0}(\Xi) = 0
\label{CS-xi}
\ee
or, for the Fourier transforms:
\be
\left( \sum_{l=1}^{n-1} p^{\mu}_{l} {\partial \over \partial p^{\mu}_{l}} 
- \omega_{J} \right) \tilde{t}_{J;K}(P;m) + \tilde{P}_{J;0}(P)= 0.
\label{CS-p}
\ee

%\newpage
\subsection{The General Case\label{general-case}}

Suppose that we have a theory with a finite number of fields, some of them of
zero-mass and some of non-zero mass. We suppose that there exists at
least one field of non-zero-mass. Then one can implement the analysis
from the case 
$m \not=0$
in the following way. Let us denote the non-zero masses of the theory as
${\bf m} \equiv (m_{1},\dots,m_{t})$;
then we have instead of (\ref{c2}) the relation:
\be
c(\lambda\lambda';{\bf m}) = 
\lambda^{\omega_{J}} c(\lambda';\lambda^{-1}{\bf m}) +
(\lambda')^{4+\omega_{k}+|\alpha|} c(\lambda;{\bf m}).
\ee

It is covenient to work in ``polar" coordinates
$(m, \mu) \in \R^{*} \times S^{t-1}$
where 
$m \equiv |{\bf m}|$
and
$\mu_{i} \equiv {m_{i}\over m}, \quad i = 1,\dots,t$.
Then, the previous relation wites as follows:
\be
c(\lambda\lambda';m,\mu) = 
\lambda^{\omega_{J}} c(\lambda';\lambda^{-1}m,\mu) +
(\lambda')^{4+\omega_{k}+|\alpha|} c(\lambda;m,\mu)
\ee
and we see that the variables
$\mu \in S^{t-1}$
play no r\^ole. Then one can implement the proof of the theorem 
\ref{m-non-zero} without any change. 
So, it follows that if at least a non-zero mass particle is present in
the theory, then the conclusions of the theorem \ref{m-non-zero} and
of the case (1) considered above for the numerical distributions are
true in this case also. More precisely, we have:
\be
{\cal U}_{\lambda}\quad  T_{J}(X;{\bf m})\quad {\cal U}_{\lambda}^{-1} =
\lambda^{\omega_{J}}\quad T_{J}(\lambda X;\lambda^{-1} {\bf m}),
\ee
\be
\left( \sum_{l=1}^{n} x^{\mu}_{l} {\partial \over \partial x^{\mu}_{l}} -
\sum_{i=1}^{t}m_{i} {\partial \over \partial m_{i}} + \omega_{J} \right) 
t_{J;K}(X;{\bf m}) = 0,
\ee
\be
\left( \sum_{l=1}^{n-1} \xi^{\mu}_{l} {\partial \over \partial \xi^{\mu}_{l}} -
\sum_{i=1}^{t}m_{i} {\partial \over \partial m_{i}} + \omega_{J}
\right) 
t_{J}(\Xi;{\bf m}) = 0
\ee
and
\be
\left( \sum_{l=1}^{n-1} p^{\mu}_{l} {\partial \over \partial p^{\mu}_{l}} +
\sum_{i=1}^{t}m_{i} {\partial \over \partial m_{i}} - \omega_{J} \right) 
\tilde{t}_{J;K}(P;{\bf m}) = 0.
\ee

Because these relations  follow from scale invariance, they can be called the 
{\it Callan-Symanzik equation} in the framework of Epstein-Glaser
perturbative scheme. However, we do not obtain the 
{\it anomalous dimension} in this way. The usual Callan-Symanzik
equation \cite{Si1}, \cite{Si2}
expresses the action of the (infinitesimal) dilation operator on the generating
functional of the Green function of the interacting field, but it is
natural to suppose that the two formalisms to be, in some way,
equivalent. We will comment more in the last Section about this point
where we will indicate the way to connect our result to the standard
arguments based on the action principle.

We close this Section with an important remark. The results contained
in the Subsections \ref{not} and in the general case here are rather
natural. Indeed, in the case fron Subsection \ref{not} of a fields of mass
$m > 0$
one could preceed more directly as follows. One can make a choice for
the chronological products
$T_{J}(X;m_{0})$
for a certain fixed mass 
$m_{0} > 0$
and {\bf define} the chronological products 
$T_{J}(X;m)$
for any other mass $m$ such that one has the simple behaviour
described by the equation (\ref{m-non-zero}). The argument can be
immediatedly adapted to the general case of more mass, but with at
least one non-zero. It is nevertheless, interesting to work out these
cases from pure cohomological considerations. So, it follows that the 
really non-trivial case is the one when all particles have null masses
and when, in principle, one cannot avoid the non-trivial cocycles of
logarithmic type.

Let us also remark that another way of proving the result from
Subsection \ref{not} is by observing that in this case one can apply
the central solution for the distribution splitting \cite{Sc1} and in
this way the scaling properties of the numerical distributions are
preserved.  
\newpage

\section{Yang-Mills Theories\label{ym}}

In this Section we analyse the scale covariance of the Standard Model (SM) and
the consequences of this property for the structure of possible anomalies. 
\subsection{The Fock Space of the Bosons}

We give some basic facts about the quantization of a spin $1$ Boson of mass
$m > 0$.
One can proceed in a rather close analogy to the case of the photon; for more
details see \cite{Gr2} and references quoted there.  Let us denote the Hilbert
space of the Boson by
${\rm H}_{m}$;
it carries the unitary representation of the orthochronous Poincar\'e group 
${\sf H}^{[m,1]}$.

The Hilbert space of the multi-Boson system should be, as before, the 
associated symmetric Fock space
${\cal F}_{m} \equiv {\cal F}^{+}({\rm H}_{m})$.
We construct this Fock space as before in the spirit of algebraic quantum 
field theory. One considers the Hilbert space 
${\cal H}^{gh}$
generated by applying on the vacuum 
$\Phi_{0}$ 
the free fields 
$A^{\mu}(x), \quad u(x), \quad \tilde{u}(x), \quad \Phi(x)$ 
which are completely characterize by the following properties:

\begin{itemize}

\item
Equation of motion:
\be
(\square + m^{2}) A^{\mu}(x), \quad (\square + m^{2}) u(x) = 0, \quad 
(\square + m^{2}) \tilde{u}(x) = 0, \quad (\square + m^{2}) \Phi(x) = 0.
\label{equ-m}
\ee

\item
Canonical (anti)commutation relations: 
\bea
[A^{\mu}(x),A^{\rho}(y)] = -g^{\mu\rho} D_{m}(x-y) \times {\bf 1}, 
\nonumber \\
~[A^{\mu}(x),u(y)] = 0, \quad [A^{\mu}(x),\tilde{u}(y)] = 0,
\quad [A^{\mu}(x),\Phi(y)] = 0,
\nonumber \\
\{u(x),u(y)\} = 0, \quad \{\tilde{u}(x),\tilde{u}(y)\} = 0, \quad
\{u(x),\tilde{u}(y)\} = D_{m}(x-y) \times {\bf 1}, 
\nonumber \\
~[\Phi(x),\Phi(y)] = D_{m}(x-y) \times {\bf 1}, \quad
[\Phi(x),u(y)] = 0.
\label{CCR-m}
\eea

\item
$SL(2,\C)$-covariance:
\bea
U_{a,A} A^{\mu}(x) U^{-1}_{a,A} = {\delta(A^{-1})^{\mu}}_{\nu}
A^{\nu}(\delta(A) \cdot x + a), 
\nonumber \\
U_{a,A} u(x) U^{-1}_{a,A} = u(\delta(A) \cdot x + a),
\quad
U_{a,A} \tilde{u}(x) U^{-1}_{a,A} = \tilde{u}(\delta(A) \cdot x + a)
\nonumber \\
U_{a,A} \Phi(x) U^{-1}_{a,A} = \Phi(\delta(A) \cdot x + a)
\label{poincare}
\eea

\item
PCT covariance.
\bea
U_{PCT} A_{\mu}(x) U_{PCT}^{-1} =  - A_{\mu}(-x), \quad
U_{PCT} \Phi(x) U_{PCT}^{-1} =  \Phi(- x)
\nonumber \\
U_{PCT} u(x) U_{PCT}^{-1} = - u(-x), \quad
U_{PCT} \tilde{u}(x) U_{PCT}^{-1} = - \tilde{u}(-x),
\nonumber \\
U_{PCT} \Phi_{0} = \Phi_{0}.
\eea

\end{itemize}

\begin{rem}
Although we could give the expressions for
$U_{I_{s}}, \quad U_{I_{t}}$
and
$U_{C}$
separately, we prefer to give only the expression of the PCT transform because
the interaction Lagrangian of the standard model is not invariant with respect
to these three operations but it is PCT-covariant.
\end{rem}

We give as before in
${\cal H}^{gh}$
the sesqui-linear form
$<\cdot,\cdot>$
which is completely characterize by requiring:
\be
A_{\mu}(x)^{\dagger} = A_{\mu}(x), \quad
u(x)^{\dagger} = u(x), \quad
\tilde{u}(x)^{\dagger} = - \tilde{u}(x), \quad
\Phi(x)^{\dagger} = \Phi(x).
\label{conjugate}
\ee

Now, the expression of the supercharge gets an extra term:
\be
Q = \int_{\R^{3}} d^{3}x  \left[ \partial^{\mu} A_{\mu}(x) + m \Phi(x) \right]
\stackrel{\leftrightarrow}{\partial_{0}}u(x) 
\label{super}
\ee
and one can see that we have
\bea
[Q, A_{\mu}] = i \partial_{\mu} u, \quad
\{Q, u \} = 0, 
\{Q, \tilde{u}\} = - i (\partial_{\mu} A^{\mu} + m \Phi) , \quad
[Q, \Phi] = i m u
\label{Q-com-m}
\eea
We still have
\be
Q^{2} = 0 \Longrightarrow Im(Q) \subset Ker(Q)
\ee
and also
\be
U_{a,A} Q = Q U_{a,A}, \quad U_{PCT} Q = - Q U_{PCT}.
\ee
Finally:
\begin{thm}
The sesqui-linear form 
$<\cdot,\cdot>$
factorizes to a well-defined scalar product on the completion of the factor
space 
$Ker(Q)/Im(Q)$.
Then there exists the following Hilbert spaces isomorphism:
\be
\overline{Ker(Q)/Im(Q)} \simeq {\cal F}_{m};
\label{factor-m}
\ee
The representation of the Poincar\'e group and the PCT operator are 
factorizing to 
$Ker(Q)/Im(Q)$
and are producing unitary operators (resp. an anti-unitary operator).
\end{thm}

If ${\cal W}$ the linear space of all Wick monomials in the fields 
$A_{\mu},~ u,~\tilde{u}$
and 
$\Phi$
acting in the Fock space
${\cal H}^{gh}$
then the expression of the BRST operator is determined by
\bea
d_{Q} u = 0, \quad d_{Q} \tilde{u} = - i (\partial^{\mu} A_{\mu} + m \Phi), 
\quad d_{Q} A_{\mu} = i \partial_{\mu} u, \quad
d_{Q} \Phi = i m u.
\label{BRSTm}
\eea
and, as a consequence we have
\be
d_{Q}^{2} = 0.
\ee

If one adds matter fields we proceed as before. In particular, this will mean
that the BRST operator acts trivially on the matter fields.

Now we can define the Yang-Mills field. We must consider the case when we have 
$r$ fields of spin $1$ and some of them will have zero mass and the others will
be considered of non-zero mass. Apparently, we need the scalar ghosts only in
the last case. However it can be shown that with this assumption, there are
no non-trivial models. To avoid this situation, we make the following
amendment. All the fields considered above will carry an  additional index
$a = 1,\dots,r$
i.e. we have the following set of fields:
$
A_{a\mu},~ u_{a},~\tilde{u}_{a},~\Phi_{a} \quad a = 1,\dots,r.
$
If one of the fields
$
A_{a\mu}
$
has zero mass we postulate that the corresponding scalar fields
$
\Phi_{a}
$
are physical fields and they will be called {\it Higgs fields}. Moreover, we do
not have to assume that they are massless i.e. if some Boson field
$A_{a}^{\mu}$
has zero mass
$m_{a} = 0$,
we can suppose that the corresponding Higgs field
$\Phi_{a}$
has a non-zero mass:
$m^{H}_{a}$.
It is convenient to use the compact notation 
\be
m^{*}_{a} \equiv \left\{\begin{array}{rcl} 
m_{a} & \mbox{for} & m_{a} \not= 0 \\
m^{H}_{a} & \mbox{for} & m_{a} = 0\end{array}\right. 
\ee

These fields verify the following equations of motion:
\bea
(\square + m_{a}^{2}) A^{\mu}_{a}(x) = 0, \quad 
(\square + m_{a}^{2}) u_{a}(x) = 0, \quad 
(\square + m_{a}^{2}) \tilde{u}_{a}(x) = 0, \quad
(\square + (m^{*}_{a})^{2}) \Phi_{a}(x) = 0 \quad
\label{equ-r}
\eea

The rest of the formalism stays unchanged.  The canonical (anti)commutation
relations are:
\bea
\left[A_{a\mu}(x),A_{b\nu}(y)\right] = - 
\delta_{ab} g_{\mu\nu} D_{m_{a}}(x-y) \times {\bf 1},
\nonumber \\
\{u_{a}(x),\tilde{u}_{b}(y)\} = \delta_{ab} D_{m_{a}}(x-y) \times {\bf 1}, 
\quad
[ \Phi_{a}(x),\Phi_{b}(y) ] = \delta_{ab} D_{m^{*}_{a}}(x-y) \times {\bf 1};
\eea
and all other (anti)commutators are null.  The supercharge is given by 
\be
Q = \sum_{a=1}^{r} \int_{\R^{3}} d^{3}x  \left[ 
\partial^{\mu} A_{a\mu}(x) + m_{a} \Phi_{a}(x) \right]
\stackrel{\leftrightarrow}{\partial_{0}}u_{a}(x) 
\label{super-r}
\ee
and verifies all the expected properties. 
                        
The Krein operator is determined by:
\be
A_{a\mu}(x)^{\dagger} = A_{a\mu}(x), \quad
u_{a}(x)^{\dagger} = u_{a}(x), \quad
\tilde{u}_{a}(x)^{\dagger} = - \tilde{u}_{a}(x), \quad
\Phi_{a}(x)^{\dagger} = \Phi_{a}(x).
\label{conjugate-YM}
\ee
                        
The ghost degree is defined in an obvious way and the expression of the BRST
operator is similar to the previous one. In particular we have:
\be
d_{Q} u_{a} = 0, \quad 
d_{Q} \tilde{u}_{a} = - i (\partial_{\mu} A_{a}^{\mu} + m_{a} \Phi_{a}) , \quad
d_{Q} A_{a}^{\mu} = i \partial^{\mu} u_{a}, \quad
d_{Q} \Phi_{a} = i m_{a} u_{a}, \quad \forall a = 1,\dots,r.
\label{BRST-YM}
\ee

Finally, the condition of gauge invariance is (see \cite{DHKS2}):
\be
d_{Q} T(X) = i \sum_{x_{l} \in X} 
{\partial \over \partial x^{\mu}_{l}} T^{\mu}_{l}(X)
\label{gauge-inv-infinites}
\ee
for some Wick polynomials
$T^{\mu}_{l}(X), \quad l = 1,\dots,|X|$.
                       
\subsection{Matter Fields and the Interaction Lagrangian of the SM}

In this case the matter field is a set of Dirac fields of mass 
$M_{A}, \quad A = 1,\dots,N$
denoted by
$\psi_{A}(x)$.

These fields are characterized by the following relations \cite{Gr3}; here
$A, B = 1, \dots, N$: 

\begin{itemize}

\item
Equation of motion:
\be
(i \gamma \cdot \partial + M_{A}) \psi_{A}(x) = 0.
\label{dirac-equ-N}
\ee

\item
Canonical (anti)commutation relations: 
\bea
[\psi_{A}(x),A_{a}^{\mu}(y)] = 0, \quad [\psi_{A}(x),u_{a}(y)] = 0, 
\quad [\psi_{A}(x),\tilde{u}_{a}(y)] = 0, \quad
[\psi_{A}(x), \Phi_{a}(y)] = 0
\nonumber \\
\{\psi_{A}(x),\psi_{B}(y)\} = 0, \quad 
\{\psi_{A}(x),\overline{\psi_{B}}(y)\} = \delta_{AB}S_{M_{A}}(x-y) 
\times {\bf 1}.
\label{CAR-m}
\eea

\item
Covariance properties with respect to the Poincar\'e group:
\be
U_{a,A} \psi_{A}(x) U^{-1}_{a,A} = S(A^{-1}) \psi_{A}(\delta(A) \cdot x + a).
\ee

\item
PCT-covariance:
\be
U_{PCT} \psi_{A}(x) U_{I_{s}}^{-1} = \gamma_{1} \gamma_{2} \gamma_{3}
\overline{\psi_{A}}(-x)^{t}.
\ee
\end{itemize}

The condition of gauge invariance remains the same (\ref{gauge-inv-infinites})
and one can prove \cite{Gr2} that this condition for
$n = 1, 2$
determines quite drastically the interaction Lagrangian of canonical dimension
$\omega(T(x)) = 4$:
\bea
T(x) \equiv
f_{abc} \left[ :A_{a\mu}(x)A_{b\nu}(x) \partial^{\nu} A_{a}^{\mu}(x): -
:A_{a}^{\mu}(x) u_{b}(x) \partial_{\mu} \tilde{u}_{c}(x):\right],
\nonumber \\
+ f'_{abc} \left[ :\Phi_{a}(x) \partial_{\mu} \Phi_{b}(x) A_{c}^{\mu}(x): 
- m_{b} :\Phi_{a}(x) A_{b\mu}(x) A_{c}^{\mu}(x): 
- m_{b} :\Phi_{a}(x) \tilde{u}_{b}(x) u_{c}(x):\right]
\nonumber \\
+ f^{"}_{abc} :\Phi_{a}(x) \Phi_{b}(x) \Phi_{c}(x): 
+  j^{\mu}_{a}(x) A_{a\mu}(x) + j_{a}(x) \Phi_{a}(x)
\label{inter}
\eea
where:
\be
j_{a}^{\mu}(x) = 
:\overline{\psi_{A}}(x) (t_{a})_{AB} \gamma^{\mu} \psi_{B}(x): +
:\overline{\psi_{A}}(x) (t'_{a})_{AB} \gamma^{\mu} \gamma_{5} \psi_{B}(x):
\label{vector-current}
\ee
and
\be
j_{a}(x) = 
:\overline{\psi_{A}}(x) (s_{a})_{AB} \psi_{B}(x): +
:\overline{\psi_{A}}(x) (s'_{a})_{AB} \gamma_{5} \psi_{B}(x):
\label{scalar-current}
\ee
are the so-called {\it currents}. The conditions of 
$SL(2,\C)$
and PCT-covariance of the interaction Lagrangian are easy to prove as well as
the causality condition. The hermiticity conditions are equivalent to 
the fact that the complex
$N \times N$
matrices
$t_{a},\quad t'_{a}, \quad s_{a}, \quad a = 1,\dots r$
are hermitian and
$s_{a}', \quad a = 1,\dots,r$
is anti-hermitian. The constants
$f_{abc}$
are completely anti-symmetric and verify Jacobi identity so they generate a
compact semi-simple Lie group quite naturally. There are other conditions on
the rest of the constants as well, but because we do not need these properties
in the subsequent analysis, we refer to the literature \cite{Gr2}, \cite{Gr3} 
and references quoted there.

Moreover, it can be proved that the condition of gauge invariance 
(\ref{gauge-inv-infinites}) is valid for 
$n = 1,2$
and we can take 
$T^{\mu}(x)$
to be of canonical dimension
$\omega(T^{\mu}(x)) = 4$
with the explicit form:
\bea
T^{\mu}(x)  = f_{abc} \left( :u_{a}(x) A_{b\nu}(x) F^{\nu\mu}_{c}(x): -
{1\over 2} :u_{a}(x) u_{b}(x) \partial^{\mu}(x) \tilde{u}_{c}(x): \right)
\nonumber \\
+ f'_{abc} \left( m_{a} :A_{a}^{\mu}(x) \Phi_{b}(x) u_{c}(x):
+ :\Phi_{a}(x) \partial^{\mu}\Phi_{b}(x) u_{c}(x): \right).
+  u_{a}(x) j^{\mu}_{a}(x).
\label{T1-YM}
\eea

The following relations are verified:
\begin{itemize}
\item
$SL(2,\C)$-covariance: for any
$A \in SL(2,\C)$
we have
\be
U_{a,A} T(x) U^{-1}_{a,A} = T(\delta(A)\cdot x+a), \quad
U_{a,A} T^{\mu}(x) U^{-1}_{a,A} = {\delta(A^{-1})^{\mu}}_{\rho}
T^{\rho}(\delta(A)\cdot x+a).
\ee
\item
PCT-covariance:
\be
U_{PCT} T(x) U^{-1}_{PCT} = T(-x), \quad
U_{PCT} T^{\mu}(x) U^{-1}_{PCT} = T^{\mu}(-x).
\ee
\item
Causality:
\be
\left[T(x), T(y)\right] = 0, \quad 
\left[T^{\mu}(x), T^{\rho}(y)\right] = 0, \quad 
\left[T^{\mu}(x), T(y)\right] = 0, \quad 
\forall x,y \in \R^{4} \quad s.t. \quad x \sim y.
\ee
\item
Unitarity:
\be
T(x)^{\dagger} = T(x), \quad
T^{\mu}(x)^{\dagger} = T^{\mu}(x).
\ee
\item
Ghost content:
\be
gh(T(x)) = 0, \quad gh(T^{\mu}(x)) = 0.
\ee
\end{itemize}

We mention that in \cite{Gr1}-\cite{Gr3}, the condition of gauge invariance is
analysed up to the order $3$.
\newpage

\subsection{Dilation Covariance of the Standard Model}

In this Subsection we generalize the arguments from the Sections \ref{dilation}
for the standard model. We denote the set of all masses by
$
{\bf m} \equiv (m_{a}, m_{a}^{*},M_{A})_{a=1.\dots,r; A=1,\dots,N}
$
and the Fock space of all particles (physical or ghosts) by
${\cal H}^{gh}_{\bf m}$.
This Hilbert space is generated from the vacuum by applying the operators:
$
A^{\mu}_{a}(x;m_{a}), \quad u_{a}(x;m_{a}), \quad 
\tilde{u}_{a}(x;m_{a}), \quad \Phi_{a}(x;m^{*}_{a})
$
and
$\psi_{A}(x;M_{A})$.
We define the dilation operators in the total Hilbert space in analogy to
(\ref{u-lambda}) and the result from the proposition \ref{dilation} stays true;
we also have the commutations relations with the Poincar\'e (\ref{reps}). 
Finally, we have from (\ref{scale-phi}) and (\ref{scale-psi}):
\bea
{\cal U}_{\lambda} A_{a}^{\mu}(x;m_{a}) {\cal U}_{\lambda}^{-1} =
\lambda A_{a}^{\mu}(\lambda x; \lambda^{-1} m_{a}), \quad
{\cal U}_{\lambda} \Phi_{a}(x;m_{a}) {\cal U}_{\lambda}^{-1}=
\lambda \Phi_{a}(\lambda x; \lambda^{-1} m_{a}),
\nonumber \\
{\cal U}_{\lambda} u_{a}(x;m_{a}) {\cal U}_{\lambda}^{-1} =
\lambda u_{a}(\lambda x; \lambda^{-1} m_{a}), \quad
{\cal U}_{\lambda} \tilde{u}(x;m_{a}) {\cal U}_{\lambda}^{-1} =
\lambda \tilde{u}(\lambda x; \lambda^{-1} m_{a}),
\nonumber \\
{\cal U}_{\lambda} \psi_{A}(x;M_{A}) {\cal U}_{\lambda}^{-1} =
\lambda^{3/2} \psi_{A}(\lambda x; \lambda^{-1} M_{A}), \quad
\forall a = 1,\dots, r, \quad \forall A = 1,\dots, N.
\eea

>From these relations and from the expressions (\ref{inter}) and (\ref{T1-YM}) 
we obtain particular cases of the relation (\ref{scale}):
\be
{\cal U}_{\lambda} T_{1}(x;{\bf m}) {\cal U}_{\lambda}^{-1} =
\lambda^{4} T_{1}(\lambda x; \lambda^{-1} {\bf m}), \quad
{\cal U}_{\lambda} T_{1}^{\mu}(x; {\bf m}) {\cal U}_{\lambda}^{-1} =
\lambda^{4} T_{1}^{\mu}(\lambda x; \lambda^{-1} {\bf m});
\label{scale-ym}
\ee
this means that both expressions have canonical dimension equal to $4$
which is also the dimension of the Minkowski space-time. Let us
suppose from now on that 
{\bf there are non-zero masses into the theory}. 
Then we can apply the
argument presented at the end of the previous Section and obtain that
for all
$|X| \geq 1$
we have the following formul\ae:
\bea
{\cal U}_{\lambda} T(X;{\bf m}) {\cal U}_{\lambda}^{-1} =
\lambda^{4|X|} T(\lambda X; \lambda^{-1} {\bf m}) 
\nonumber \\
{\cal U}_{\lambda} T^{\mu}_{l}(X;{\bf m}) {\cal U}_{\lambda}^{-1} =
\lambda^{4|X|} T^{\mu}_{l}(\lambda X; \lambda^{-1} {\bf m}).
\label{scale-ym-n}
\eea

We also mention the following result which easily follows from the definitions:
\begin{lemma}
The following relations is valid for every Wick monomial:
\be
{\cal U}_{\lambda} \left[ d_{Q} W(X; {\bf m})\right] {\cal U}_{\lambda}^{-1} =
\lambda^{\omega(W)+1} W(\lambda X; \lambda^{-1} {\bf m}).
\label{dQ-lambda}
\ee
\end{lemma}

{\bf Proof:}
If the expression $W$ is one of the fields
$
A^{\mu}_{a}(x;m_{a}), u_{a}(x;m_{a}), \tilde{u}_{a}(x;m_{a}), 
\Phi_{a}(x;m^{*}_{a})$
or
$\psi_{A}(x;M_{A}), \bar{\psi}_{A}(x;M_{A})$
the formula from the statement follows elementary; then we extend to any Wick
monomial by induction, using the derivative properties of the BRST operator.
$\qed$

%\newpage
\subsection{The Structure of the Anomalies in the Standard Model}

We consider the standard model as defined by the Lagrangian (\ref{inter}). 
and suppose that there are no anomalies up to the order 
$n - 1$
i.e. we have (\ref{gauge-inv-infinites}) up to this order. The purpose of this
Subsection is to find if possible anomalous terms can appear in this relation
in order $n$ and what limitation are imposed by scale covariance. The analysis
is similar to the case of the quantum electrodynamics
\cite{Gr5}. However, we prefer to use the formalism developped in
Subsections \ref{bogoliubov} and \ref{EG}.

(i) Suppose that we have constructed the chronological products
$T_{J}(X), \quad |X| \leq n - 1$
verifying all the induction hypothesis from Subsection \ref{EG}.
Then we will be able to use the formul\ae~of the type 
(\ref{average-chrono}).   

We must have, in analogy to (\ref{one}), a developpment of the type:
\be
T^{\mu}(x) = \sum c^{\mu}_{j} T_{j}(x)
\label{one-mu}
\ee
with 
$c^{\mu}_{j}$
some real constants; then we will have in analogy to (\ref{one-n}):
\be
T^{\mu}_{l}(X) = \sum c_{j_{1}} \dots c^{\mu}_{j_{l}} \dots c_{j_{n}} 
T_{j_{1},\dots,j_{n}}(X)
\label{one-n-mu}
\ee
for all
$|X| \leq n - 1$.
In particular, the following conventions hold:
\be
T(\emptyset) \equiv {\bf 1}, \quad 
T^{\mu}_{l} (\emptyset) \equiv 0, \quad
T^{\mu}_{l} (X) \equiv 0, \quad {\rm for} \quad x_{l} \not\in X.
\label{empty-p}
\ee

We supplement the induction hypothesis adding: 
\begin{itemize}
\item
ghost number content:
\be
gh(T(X)) = 0, \quad gh(T^{\mu}_{l}(X)) = 1, \quad |X| \leq n-1;
\label{gh-p-ym}
\ee
\item
gauge invariance:
\be
d_{Q} T(X) = i \sum_{l=1}^{n} {\partial \over \partial x^{\mu}_{l}} 
T^{\mu}_{l}(X), \quad |X| \leq n-1.
\label{gauge-p}
\ee
\item
scale covariance:
\be
{\cal U}_{\lambda} T_{J}(X;{\bf m}) {\cal U}_{\lambda}^{-1} =
\lambda^{\omega_{J}} T_{J}(\lambda X; \lambda^{-1} {\bf m}), 
\quad |X| \leq n-1.
\label{scale-p}
\ee
\end{itemize}

(ii) Now we can construct the expressions
$D_{J}(X;{\bf m})$
according to the formula (\ref{com-D}) from Subsection \ref{EG} such 
that we have the well known properties of causality, Poincar\'e 
covariance and unitarity. We consider now a causal splitting of the
type (\ref{decD}) such that we preserve Poincar\'e covariance and the
order of singularity. The chronological products can be obtained from
the formula (\ref{chronos-n}), but we still have some freedom in the
choice of the splitting which will shall use in the following.
It can be proved as in \cite{Gr5} that we have, instead of the relation
(\ref{gauge-p}) a somewhat weaker form, namely:
\be
d_{Q} A(X;{\bf m}) = i \sum_{l=1}^{n} {\partial \over \partial x^{\mu}_{l}} 
A^{\mu}_{l}(X;{\bf m}) + P(X;{\bf m}), \quad |X| = n
\label{anoP}
\ee
where the expressions
$A(X;{\bf m})$
and
$A^{\mu}_{l}(X;{\bf m})$
are constructed from the expressions 
$A_{J}(X)$
according to the prescriptions (\ref{one-n}) and (\ref{one-n-mu}). In the
right hand side
$P(X;{\bf m})$
is a Wick polynomial (called {\it anomaly}) of the following structure:
\be
P(X) = \sum_{J} \left[ p_{J}(\partial) \delta(X) \right]
:T_{j_{1}}(x_{1}) \cdots T_{j_{n}}(x_{n}):
\label{wickP}
\ee
with
$p_{J}$
polynomials in the derivatives with the maximal degree restricted by
\be
deg(p_{J}) + \omega_{J} \leq 5, \quad \forall J.
\label{degP}
\ee

If we argue like in theorem \ref{m-non-zero} then we can see that from
the induction hypothesis we have the following scaling behaviour of
the chronological products
$T_{J}(X), \quad |X| = n$:
\be
{\cal U}_{\lambda} T_{J}(X;{\bf m}) {\cal U}_{\lambda}^{-1} =
\lambda^{\omega_{J}} T_{J}(\lambda X; \lambda^{-1} {\bf m}) 
+ P_{J;k;{\bf m};\lambda}(X) T_{k}(x_{n};{\bf m}) 
\label{ano-scale}
\ee
for some quasi-local distribution
$P_{J;k;{\bf m};\lambda}(X)$
having an expression of the form (\ref{p-delta}).
Moreover, these distributions have a coboundary structure (because we
are in the case when we have massive fields in the theory) so one can
redefine the chronological products
$T_{J}(X)$
such that one gets rid of the expression
$P_{J;k;{\bf m};\lambda}(X)$.
This implies, redefinitions for the causal splitting 
(\ref{decD}), in particular redefinitions for the distributions
$A(X;{\bf m})$
and
$A^{\mu}_{l}(X;{\bf m})$. 
It is clear that in this way we will not affect the general structure
of the equation (\ref{anoP}), that is we eventually modify the anomaly
$P$ without spoiling the Poincar\'e covariance and the order of
singularity of the splitting. Moreover, we will have in this way,
instead of (\ref{ano-scale}), the relation (\ref{scale-p}) for 
$|X| = n$
also.
Because we can prove from the induction hypothesis that we have
\be
{\cal U}_{\lambda} A'_{J}(X;{\bf m}) {\cal U}_{\lambda}^{-1} =
\lambda^{\omega_{J}} A'_{J}(\lambda X; \lambda^{-1} {\bf m}) 
\ee
for
$|X| = n$,
we obtain that in this case we have also
\be
{\cal U}_{\lambda} A_{J}(X;{\bf m}) {\cal U}_{\lambda}^{-1} =
\lambda^{\omega_{J}} A_{J}(\lambda X; \lambda^{-1} {\bf m}) 
\ee
and in particular:
\be
{\cal U}_{\lambda} A(X;{\bf m}) {\cal U}_{\lambda}^{-1} =
\lambda^{4n} A(\lambda X; \lambda^{-1} {\bf m}),
\quad
{\cal U}_{\lambda} A^{\mu}_{l}(X;{\bf m}) {\cal U}_{\lambda}^{-1} =
\lambda^{4n} A^{\mu}_{l}(\lambda X; \lambda^{-1} {\bf m}).
\label{scale-A}
\ee

This result can be obtained combining with the gauge invariance condition
given by the equation (\ref{anoP}) in the following way: we apply to 
${\cal U}_{\lambda} \cdots {\cal U}_{\lambda}^{-1}$
to (\ref{anoP}), we use the lemma \ref{dQ-lambda} and then the
previous relations. The result is the following identity verified by
the 
anomaly:
\be
{\cal U}_{\lambda} P(X;{\bf m}) {\cal U}_{\lambda}^{-1} =
\lambda^{4n+1} P(\lambda X; \lambda^{-1} {\bf m}).
\label{scale-ano}
\ee

In other words, the anomaly must have the canonical dimension equal to
$4n + 1$.
By ``integrations by parts" (see \cite{Gr5}) we can exhibit the anomaly as
follows:
\be
P(X) =  i \sum_{l=1}^{n} {\partial \over \partial x^{\rho}_{l}}
N^{\rho}_{l}(X) + P'(X)
\label{int-parts}
\ee
where
$
P'(X)
$
is of the following form:
\be
P'(X) = \delta(X) {\cal P}(x_{n})
\ee
with 
$
{\cal P}(x)
$
a Wick polynomial in one variable. So, by redefining the expressions
$A^{\mu}_{l}(X)$
we can take the anomaly of the form
\be
P(X) = \delta(X) {\cal P}(x_{n}).
\label{generic}
\ee

It is obvious that the ``integration by parts" process will not affect
the properties of the anomaly that we have already obtained. In
consequence, the Wick polynomial 
$
{\cal P}(x)
$
will verify the following restrictions:
\begin{itemize}
\item
$SL(2,\C)$-covariance:
\be
U_{a, A} {\cal P}(x) U^{-1}_{a, A} = {\cal P}(\delta(A)\cdot x + a), 
\quad \forall (a,A) \in inSL(2,\C).
\label{invariance-P-cal}
\ee
\item
Ghost numbers restrictions:
\be
gh({\cal P}(x)) = 1.
\label{ghP-ym-cal}
\ee
\item
Scale covariance:
\be
{\cal U}_{\lambda} {\cal P}(x;{\bf m}) {\cal U}_{\lambda}^{-1} =
\lambda^{5} {\cal P}(\lambda X; \lambda^{-1} {\bf m}).
\label{dimension}
\ee
\end{itemize}

If we use the generic structure
\be
{\cal P}(x;{\bf m}) = \sum_{j} c_{j}({\bf m}) \quad T_{j}(x;{\bf m})
\ee
with
$c_{j}({\bf m})$
some mass-dependent constants, in the last equation we immediately get:
\be
c_{j}(\lambda {\bf m})  = \lambda^{5-\omega_{j}} c_{j}({\bf m}).
\label{scale-ano-omega}
\ee

If we now construct the chronological products 
$T(X), \quad T^{\mu}_{l}(X)$
one can fix in a standard way (see \cite{EG1}) the properties of 
symmetrization and unitarity without spoiling the other relations we
have already obtained. In the same one can fix PCT-covariance. In the
end, we will have a relation of the type:
\be
d_{Q} T(X;{\bf m}) = i \sum_{l=1}^{n} {\partial \over \partial x^{\mu}_{l}} 
T^{\mu}_{l}(X;{\bf m}) + P(X;{\bf m}), \quad |X| = n
\label{ano-chrono}
\ee
where the anomaly
$P(X,{\bf m})$
has the form (\ref{generic}) and verifies all the preceding restrictions: 
Poincar\'e covariance, ghost number restriction, scale covariance.
Moreover, it obviously verifies gauge invariance:
\be
d_{Q} P(X;{\bf m}) = i \sum_{l=1}^{n} {\partial \over \partial x^{\mu}_{l}} 
P^{\mu}_{l}(X;{\bf m})
\ee
for some Wick polynomials
$P^{\mu}_{l}(X;{\bf m})$
and also PCT-covariance:
\be
U_{PCT} {\cal P}(x) U^{-1}_{PCT} = (-1)^{n} {\cal P}(-x)
\label{PCT-cal}
\ee
and unitarity:
\be
{\cal P}(x)^{\dagger} \equiv (-1)^{n} {\cal P}(x).
\label{uniP-ym-cal}
\ee

(iii) The list of possible anomalies can be written now as in \cite{Gr5}. We
only remark that the restrictions imposed above do not lead to the conclusion
that there are no anomalies in order $n$. In fact, a number of {\it hard
anomalies} remain such as:

\be
{\cal P}_{1} = c^{1}_{abcde} \sum_{m_{a} = m_{b} = m_{c} = m_{d} = m_{e} = 0}
u_{a} :\Phi_{b} \Phi_{c} \Phi_{d} \Phi_{e}:
\ee

\be
{\cal P}_{2} = c^{2}_{abc} \sum_{m_{a} = m_{b} = m_{c} = 0}
u_{a} :\partial^{\mu}\Phi_{b} \partial_{\mu}\Phi_{c}: 
\ee

\be
{\cal P}_{3} = c^{3}_{abc}  \varepsilon_{\mu\nu\rho\sigma} 
u_{a} :F_{b}^{\mu\nu} F_{c}^{\sigma\rho}: 
\label{ABBJ}
\ee
where
\be
F_{a}^{\mu\nu} \equiv 
\partial^{\mu} A^{\nu}_{a} - \partial^{\nu} A^{\mu}_{a}.
\ee

\be
{\cal P}_{4} = \sum_{m_{a} = m_{b} = 0} \left[
:\bar{\psi}_{A} (K_{ab})_{AB} \psi_{B}: 
:\bar{\psi}_{A} (K'_{ab})_{AB} \gamma_{5} \psi_{B}: \right] 
u_{a} \Phi_{b}.
\ee

One can show that from unitarity (or PCT-covariance) that we have

\be
c_{\dots} = (-1)^{n} c_{\dots}, \quad
K_{ab}^{*} = (-1)^{n} K_{ab}, \quad (K'_{ab})^{*} = (-1)^{n} K'_{ab}.
\ee

The list of hard anomalies is larger: all the anomalies appearing in the second
and in the third order of perturbation theory (see \cite{Gr2} and \cite{Gr3})
should appear. 

\section {Conclusions}

The expression (\ref{ABBJ}) is the famous Adler-Bardeen-Bell-Jackiw anomaly
(ABBJ). So, we see that the various symmetries of the standard model (including
scale covariance) are not sufficient to prove the anomalies are absent in
higher orders of the perturbation theory if they are absent in orders 
$n = 1,2,3$ 
(at least in Epstein-Glaser approach). In fact, if a certain type of anomaly is
present in low orders of perturbation theory, this means that the corresponding
expression is not in conflict with the various symmetries of the model. Then it
is hard to imagine why such a conflict would appear in higher orders of
perturbation theory. Such a result would be possible in our formalism only if
in the equation (\ref{scale-ano-omega})
the number $n$ (the order of the perturbation theory) would survive.

To obtain the cancelation of anomalies in all orders in our formalism
a more refined formula for the distribution splitting seems to be needed. 

There appears to be a contradiction between our result and 
the analysis from \cite{BBBC} (see also
\cite{BRS} and \cite{PR}) where it is showed that the ABBJ anomaly can appear
only in the order
$n = 3$.

The discrepancy can be explained if one admits that in these
references one works with the {\bf interaction fields}. One know that
one can construct such fields from the chronological products
$T_{J}(X)$
as formal series ( see formula (76) of \cite{EG1}) of the type
\be
\Phi(x) = \sum {i^{n} \over n!} \int dy_{1} \dots dy_{n}
R(y_{1},\dots,y_{n};x) g(y_{1}) \cdots g(y_{n})
\label{int-field}
\ee
where 
$R$ are the retarded products. If we perform formally the adiabatic
limit we have
\be
\Phi(x) = \sum_{n=0}^{\infty} g^{n}~\phi_{n}(x)
\label{int-field1}
\ee 
where
$\phi_{0}$
is the free field and $g$ is the coupling constant.

Now  supposes that a formula of the following type is valid:
\be
{\cal U}_{\lambda}\quad  \Phi(x)\quad {\cal U}_{\lambda}^{-1} =
\lambda^{\omega - \gamma(g)}\quad \Phi(\lambda x),
\label{ano-dim}
\ee
where
$\omega$
is the canonical dimension of the field 
$\phi$
and
\be
\gamma(g) = \sum_{n=1}^{\infty} \gamma_{n} g^{n}
\ee
is a formal series in the coupling constant called 
{\it anomalous dimension}. One can write 
\be
\lambda^{-\gamma(g)} = e^{-\gamma(g) ln(\lambda)}
\ee
perform the Taylor expansion in $g$ and substitute in the preceding
formula. Then one finds out by regrouping the terms that we have
\be
{\cal U}_{\lambda}\quad  \phi_{n}(x)\quad {\cal U}_{\lambda}^{-1} =
\lambda^{\omega}\quad \sum_{p+q=n} \sum_{m_{1},\dots,m_{p}}
{(- ln\lambda)^{m_{1} + \cdots m_{p}} \over p!} 
\gamma_{m_{1}}\cdots \gamma_{m_{p}} \Phi_{q}(\lambda x).
\label{ano-dim-R}
\ee

This relation should be interpreted as a relation on the retarded
products $R$. In fact such a relation is compatible with our analysis
in the scaling limit, where all momenta are very large. In this region
it is plausible to assume that all masses of the theory are zero, so
we can apply the result of Subsection \ref{zero} which leads to a
formula having the structure (\ref{ano-dim-R}). 

It is an interesting problem to make this analysis completely rigourous
in the framework of Epstein-Glaser formalism.
\vskip 1cm
{\bf Acknowledgement}: The author wishes to thank prof. K. Fredenhagen
and dr. M. D\"utsch for many discussions and critical remarks. 

\end{document}